\documentclass[varenna]{cimento}
\title{Theory of excitations of the condensate \\ and non-condensate
at finite temperatures }
\shorttitle{Theory of Excitations at Finite Temperatures}
\author{A. Griffin}
\institute{Department of Physics, University of Toronto, 
Toronto, Ontario M5S 1A7, Canada} 
\PACSes{\PACSit{03.75.Fi, 05.30.Jp, 67.40.-w.}{}}

\begin{document}
\maketitle
\begin{abstract}
We give an overview of the current theory of collective modes in 
trapped atomic gases at finite temperatures, when the dynamics of the 
condensate and non-condensate must both be considered.  A simple 
introduction is given to the quantum field formulation of the 
dynamics of an interacting Bose-condensed system, based on equations 
of motion for the condensate wavefunction and single-particle Green's 
functions for the non-condensate atoms.  We discuss the nature of 
excitations in the mean-field collisionless region, including the 
Beliaev second-order approximation for the self-energies.  We also 
sketch the derivation of coupled two-fluid hydrodynamic equations 
using a simple kinetic equation which includes collisions between 
condensate and non-condensate atoms.

\end{abstract}
\section{Introduction}
\label{sec:Introduction}

There are several excellent review articles on trapped Bose-condensed 
atomic gases written at a relatively introductory level 
\cite{ref:Dalgiopitstr,ref:Pethsmi}, 
with emphasis on the dynamics of the condensate at $T=0$.
In the present lectures, 
I will concentrate on the interplay between the condensate 
and non-condensate components.
This topic requires a more sophisticated analysis based 
on the concepts and methods of many body theory.
My lectures will attempt to give a very basic introduction 
to this kind of approach.
The primary audience I have in mind are graduate students 
and postdocs coming from atomic and laser physics, 
rather than condensed matter theory.
However, I hope the experts will also learn something from these 
lectures.
In his lectures, Fetter \cite{ref:a} has given a detailed review of 
excitations in a trapped dilute Bose gas at $T=0$.
He shows that a convenient way of discussing these excitations is 
to start from the time-dependent Gross-Pitaevskii (GP) equation for 
the 
macroscopic wavefunction $\Phi({\bf r},t)$.  
Linearizing around the equilibrium value $\Phi_0 ({\bf r})$, 
one finds 
 of the macroscopic wavefunction are given by 
the well-known Bogoliubov coupled equations.
For a uniform Bose-condensed gas, the excitation frequencies are 
those first 
discussed by Bogoliubov in 1947 \cite{ref:b}.

At $T=0$, one can assume that all the atoms are in the Bose 
condensate described by 
$\Phi({\bf r},t)$.  
In contrast, in these lectures, my main topic will be to review our 
understanding of what excitations are in a Bose-condensed gas at {\it 
finite} temperatures, when there is a large number of atoms in the 
non-condensate (in trapped gases, the non-condensate is often 
referred to as the ``thermal cloud'').
I will make some contact with the ideas discussed in the lectures by 
Burnett \cite{ref:Bur}.

The first half of these lectures (Sections~\ref{sec:Elementary 
excitations and density fluctuations in normal systems} - 
\ref{sec:Why are excitations in a Bose fluid so interesting}) deals 
with excitations whose very existence depends on self-consistent mean 
fields (of various kinds!), rather than on the  collisions between 
atoms.
In the standard language developed in condensed matter physics in the 
1960's, this means the excitations are in the ``collisionless 
region''.

The second half of the lectures (Sections~\ref{sec:Hydrodynamic 
oscillations in a trapped Bose gas} - \ref{sec:First and second sound 
in a uniform Bose-condensed gas}) deals with excitations in the 
collision-dominated hydrodynamic region.
I  review the two-fluid hydrodynamic equations such as given by 
Landau \cite{ref:Land}, generalized to include a trap.
I  give an explicit microscopic derivation of such two-fluid 
equations in a trapped Bose gas.   This extends recent work of 
Zaremba, Griffin and Nikuni (ZGN) \cite{ref:Zargrinik} to include the 
case when the condensate is not yet in local equilibrium with the 
non-condensate atoms.

The type of questions I want to address in these lectures include:

\begin{enumerate}
\item What is the difference between an elementary excitation and a 
collective mode?

\item Can we {\it isolate} the dynamical role of the condensate on 
the nature of the excitations?

\item At $T=0$, with a pure condensate, excitations in a gas must be 
oscillations of the condensate.
In contrast, above $T_{\rm BEC}$, 
the excitations are not related to a condensate.
What happens to an excitation as we go from $T=0$ to  $T > T_{\rm 
BEC}$. How does the excitation get rid of its ``condensate'' dressing?

\item What is the essential physics behind the different mean-field 
theories of excitations which have been discussed in the recent 
literature:  Gross-Pitaevskii, Hartree-Fock-Popov, 
Hartree-Fock-Bogoluibov?
\item What is the physics behind the dreaded Beliaev second-order 
approximation?  This is the first approximation which includes 
damping of the elementary excitations even at $T=0$ ~\cite{ref:Bel1}.
\item Why are the excitations and collective modes in a 
Bose-condensed system uniquely interesting, compared to all other 
many body systems?  The key reason is, of course, that the condensate 
couples and hybridizes single-particle excitations with density 
fluctuations.
Above $T_{\rm BEC}$, these two excitation branches are uncoupled.
Once this is understood, one sees why more detailed and systematic 
experimental studies of excitations in trapped atomic gases at finite 
temperature are of great importance.

\end{enumerate}
To give a careful discussion of all these questions would need~10 
lectures.
In these~3 lectures, I will give a speeded-up version.  
I will often use a {\it uniform} weakly interacting Bose gas to 
illustrate the structure of the theory.
While I will only sketch the math, I will still try to give a flavour 
of what is involved.
The approach I will use to discuss these questions is based on the 
field-theoretic formulation of a Bose-condensed system of particles.
As I review elsewhere in this volume~\cite{ref:Grif1}, this powerful 
formalism was introduced by Beliaev~\cite{ref:Bel2} in 1957 and 
extensively developed in the Golden Period: 1958-1965.
I will introduce this formalism in a very schematic manner - but even 
experimentalists will find it useful to know some of the ``language" 
used in this approach.
There are other methods to deal with collective modes in Bose gases 
but they are not as useful at {\it isolating} the dynamical role of 
the condensate, or dealing with finite temperatures.

While I will always have trapped atomic Bose gases in mind, much of 
the general theory~\cite{ref:Grif2, ref:Fetwal} is valid for any 
Bose-condensed fluid (gas or liquid).
Thus I will often make references to superfluid $^4$He, pointing out 
similarities with Bose gases.

\section{Elementary excitations and density fluctuations in normal 
systems}
\label{sec:Elementary excitations and density fluctuations in normal 
systems}

We work with quantum field operators :
\begin{eqnarray}
{\hat\psi}({\bf r}) & =& \mbox{destroys an atom at}\ {\bf 
r}\nonumber\\
{\hat \psi}^+({\bf r}) &=& \mbox{creates an atom at}\  {\bf 
r}.\label{eq:elementary excitations1}\end{eqnarray}  
These 
operators satisfy the usual Bose commutation relations, 
such as $[\hat{\psi}({\bf r}),\hat{\psi}^+({\bf r'})]=\delta ({\bf r} 
- {\bf r'})$.  Of course, if we have several hyperfine atomic states, 
then we have different field operators $\hat{\psi}_a({\bf r})$, 
where $a$ is the hyperfine state label (the analogue of a spin 
label).  All observables can be written in terms of these quantum 
field operators.  Two important examples are the local density 
operator:
\begin{equation}
{\hat n}({\bf r}) = {\hat\psi}^+({\bf r}){\hat\psi}({\bf r})
\label{eq:elementary excitations2}\end{equation}
and the Hamiltonian
\begin{equation}
{\hat H} = \int d{\bf r} {\hat\psi}^+({\bf r})\left[{-\nabla^2 \over 
2m} + U_{\rm ext}({\bf r})-\mu\right] {\hat\psi}({\bf r}) + {1 \over 
2} g \int d{\bf r} \hat{\psi}^+({\bf r}) \hat{\psi}^+({\bf r}) 
\hat{\psi}({\bf r}) \hat{\psi}({\bf r}).
\label{eq:elementary excitations3}\end{equation}
In these lectures, the two-particle interaction will always be 
approximated 
by a $s$-wave scattering length, 
appropriate to a dilute Bose gas at very low temperatures 
\cite{ref:Bur, ref:Heindali}.
Thus $v ({\bf r}-{\bf r'}) = g \delta({\bf r}-{\bf r'})$, with $g 
={4\pi a\over m}$ (we set $\hbar =1$ throughout this article).

What we want to calculate are various kinds of correlation functions 
involving different local operators:
\begin{eqnarray}
\langle{\hat n} ({\bf r},t) {\hat n}({\bf r'}, t')\rangle&\sim& 
\chi_{nn}({\bf r}t,{\bf r'}t')\nonumber \\
&=&\chi_{nn}(1,1') \equiv \mbox{density response function}\nonumber \\
\langle {\hat j}_\alpha({\bf  r},t){\hat j}_\beta({\bf r'}t')\rangle 
&\sim& \chi_{j_\alpha j_\beta} ({\bf  r}t, {\bf  r'}t')\nonumber\\ 
&=& \chi_{ j_\alpha j_\beta}(1,1') \equiv\mbox{current response 
function}.
\label{eq:elementary excitations4}\end{eqnarray}
In these tutorial lectures, I will not bother to distinguish between 
the various kinds of correlation functions (time-ordered, retarded, 
etc) or even the difference between correlation functions and 
response functions.
These are discussed in all the standard textbooks on many body theory 
\cite{ref:Fetwal,ref:Pinesnoz}
but are only important when one is doing systematic calculations.
A convenient summary of the formalism is given in a recent review 
article on homogeneous weakly interacting Bose gases 
\cite{ref:Shigrif}.

The density response function
\begin{equation}\chi_{nn}(1,1') = \langle 
{\hat{\psi}^+}(1){\hat\psi}(1){\hat\psi}^+(1'){\hat\psi}(1')\rangle 
\label{eq:elementary excitations5}
\end{equation}
involves four quantum field operators and is an example of a 
two-particle Green's function $G_2(1,1')$.  One can measure 
$\chi_{nn} (1,1')$ by coupling a weak  external field to the local 
density in the system
\begin{equation}
H'(t) = \int d{\bf r} V_{\rm ex}({\bf r}, t){\hat n}({\bf 
r})\equiv\int d 1 V_{\rm ex}(1){\hat n}(1).\label{eq:elementary 
excitations6}
\end{equation}
Linear response theory gives
\begin{eqnarray}
\delta n(1)&\equiv&\langle{\hat n}({\bf r})\rangle_t -\langle{\hat 
n}({\bf r})\rangle_{eq}\nonumber \\
&=& \int d1'\chi_{nn}(1, 1') V_{\rm ex}(1') + \dots
\label{eq:elementary excitations7}
\end{eqnarray}
for the density response (see Chapters 5 and 9 of 
Ref.\cite{ref:Fetwal} and Chapter 2 of Ref.\cite{ref:Pinesnoz}).
For a uniform system with $V_{\rm ex}({\bf r}, t) = V_{q,\omega}\  
e^{i({\bf q}\cdot{\bf  r}-\omega t)}$, we have
\begin{eqnarray*}
\chi_{nn} (1, 1') &=& \chi_{nn} (1 - 1') \\
&=& \chi_{nn}({\bf r} - {\bf r'}, t - t').\end{eqnarray*}
Fourier transforming $({\bf r}-{\bf r'})\rightarrow {\bf q}$ and $(t 
- t')\rightarrow\omega$, the linear response equation in 
(\ref{eq:elementary excitations7}) reduces to
\begin{equation}
\delta n({\bf q}, \omega)=\chi_{nn}({\bf q}, \omega)V_{q, 
\omega}.\label{eq:elementary excitations8}
\end{equation}
If $\chi_{nn}({\bf q}, \omega) \sim {1\over \omega - E_q}$ has a pole 
at $\omega = E_q$, (for further details, see Section~\ref{sec:Density 
fluctuation spectrum in the mean-field approximation}) then when 
$\omega$ and $q$ of the external potential satisfy $\omega = E_q$, 
 $\delta n({\bf q}, \omega)$ as given by (\ref{eq:elementary 
excitations8})
can be very large even though $V_{q, \omega}$ is small.
We note that
\begin{equation}
\chi_{nn}({\bf q}, \omega) \sim {1\over \omega - E_q} \rightarrow 
\chi_{nn}({\bf q}, t) \sim e^{-iE_qt}\label{eq:elementary 
excitations9}
\end{equation}
and thus clearly the pole at $\omega = E_q$ is the signature of an 
oscillating density fluctuation.

The basic correlation function in the field-theoretic approach is 
given by
\begin{equation}
\langle{\hat\psi}(1){\hat\psi}^+ (1')\rangle \sim G_1(1, 
1').\label{eq:elementary excitations10}
\end{equation}
This single-particle Green's function $G_1$ involves two quantum 
field operators.
It describes creating an atom at $1' = {\bf r'}, t'$, let it 
propagate through the system to $1 = {\bf r}, t$ and then destroying 
the atom.
All other higher-order correlation functions such as $G_2(1, 1')$ can 
be constructed out of combinations of $G_1(1, 1')$.
For example, the {\it lowest order} contributions to the density 
response function in (\ref{eq:elementary excitations5}) are
\begin{eqnarray}
\chi_{nn} (1,1') &=& 
\langle{\hat\psi}^+(1){\hat\psi}(1){\hat\psi}^+(1'){\hat\psi}(1')\rangle 
\nonumber\\
&\simeq &\langle{\hat\psi}^+(1){\hat\psi}(1)\rangle 
\langle{\hat\psi}^+(1'){\hat\psi}(1')\rangle \rightarrow \langle 
n(1)\rangle \langle n(1')\rangle \nonumber\\
&+ &\langle {\hat\psi}^+(1){\hat\psi}(1')\rangle 
\langle{\hat\psi}^+(1')\psi(1)\rangle \rightarrow G_1(1, 1') G_1(1', 
1) \nonumber\\
&+& \langle {\hat\psi}^+(1){\hat\psi}^+(1')\rangle 
\langle{\hat\psi}(1) {\hat\psi} (1') \rangle + \dots
\label{eq:elementary excitations11}
\end{eqnarray}
As we discuss in Section~\ref{sec:Green's function formulation},  the 
terms in the last line of (\ref{eq:elementary excitations11}) vanish 
in a normal Bose system but are finite for $T<T_{\rm BEC}$.
The poles of $G_1$ correspond to what are called elementary 
excitations (or quasiparticles).
In a uniform system, we have
\begin{eqnarray}
& & G_1(1, 1') \rightarrow G_1({\bf q}, \omega) \sim {1\over \omega - 
E_q^{sp}} \nonumber\\
& &G_1 ({\bf q}, t) \sim e^{-iE_q^{sp}t};\  \mbox{in a free gas, we 
have}\  E_q^{sp} = {q^2 \over 2m}\equiv 
\epsilon_q.\label{eq:elementary excitations12}
\end{eqnarray}
One can show \cite{ref:Fetwal} that these single-particle excitations 
determine the thermodynamic properties of interacting systems.
However, it is very difficult to directly measure the spectrum of 
$G_1(1, 1')$ since one needs an external  field which couples to 
${\hat\psi}(1)$, ie, an atom reservoir.
 Later we will see that what makes a Bose-condensed system unique is 
that we can easily access $G_1(1, 1')$ as a result of the effects of 
the Bose condensate.

Finally we introduce the key idea of a single-particle self-energy 
through Dyson's equation:
\begin{equation}
G_1 = G_0 + G_0 \Sigma G_1,\label{eq:elementary 
excitations13}\end{equation}
where $G_1$ is the interacting single-particle Green's function, 
$G_0$ is the non-interacting single-particle Green's function
and all effects of the two-particle interactions are contained in the 
self-energy function $\Sigma  $.
In a uniform system, we can Fourier transform this Dyson equation to 
give
\begin{equation}
G_0(q, \omega) = G_0(q, \omega) + G_0 (q, \omega) \Sigma (q, \omega) 
G_1 (q, \omega),
\label{eq:elementary excitations14}\end{equation}
where $G_0(q, \omega) = {1 \over \omega-\epsilon_q}$.  This is easily 
solved to give
\begin{equation}
G_1(q, \omega) = {1 \over \omega - [\epsilon_q + \Sigma(q, 
\omega)]}.\label{eq:elementary excitations15}
\end{equation}
Thus we see that $G_1$ may have a single-particle pole at the 
quasiparticle energy
\begin{equation} E_q^{sp} = \epsilon_q +\Sigma (q, 
E_q^{sp}).\label{eq:elementary excitations16}\end{equation}
In general, $\Sigma(q, \omega) = \Sigma_R + i\Sigma_I$, where 
$\Sigma_I $ describes the damping of the single-particle excitations.

Field-theoretic calculations \cite{ref:Grif2,ref:Fetwal}
involve a systemmatic (diagrammatic) procedure to calculate 
$\Sigma(q, \omega)$ and from this to obtain $G_1(q, \omega)$.
We note that the self-energy $\Sigma$ is by definition highly 
non-perturbative.
As an illustration, let us consider the self-energy to first order in 
$g$.
This Hartree-Fock approximation is shown in Fig.1.
In our $s$-wave approximation, the total self-energy is simply 
  
\begin{equation}
\Sigma_{HF} = gn + gn = 2gn, \label{eq:elementary excitations17} 
\end{equation}
and therefore (\ref{eq:elementary excitations15}) gives
\begin{equation}
G({\bf q}, \omega) = {1 \over \omega-[\epsilon_q +2gn]}.
\label{eq:elementary excitations18}
\end{equation}
Hence the normal HF excitation energy has the dispersion relation
\begin{equation}
\ E_q^{HF} = {q^2\over 2m}+2gn.\label{eq:elementary excitations19}
\end{equation}

\section{Density fluctuation spectrum in the mean-field approximation}
\label{sec:Density fluctuation spectrum in the mean-field 
approximation}

The simplest approximation for the density response function 
introduced in Section~\ref{sec:Elementary excitations and density 
fluctuations in normal systems}
 is \cite{ref:Grif2,ref:Fetwal}
\begin{equation}
\chi_{nn}(1, 1') \Rightarrow \chi^0_{nn} (1, 1') = G_1 (1, 1') G_1 
(1', 1).
\label{eq:density fluc1}
\end{equation}
In a uniform Bose system, the Fourier transform of this gives
\begin{equation}
\chi^0_{nn}(q, \omega)\sim\int d{\bf k}\int d\omega_1 \int d\omega_2 
{\left[N^0(\omega_1)-N^0(\omega_2)\right]\over
\omega_1 - \omega_2 - \omega} A({\bf k}, \omega_1) A({\bf k} - {\bf 
q}, \omega_2).
\label{eq:density fluc2}
\end{equation}
Here $N^0(\omega) = (e^{\beta\omega}-1)^{-1}$ is the Bose 
distribution function and
$A({\bf k}, \omega) \equiv $ single-particle spectral density $\sim 
2{\rm Im} G_1 ({\bf k}, \omega +i0^+)$. If we use $A_{HF}(k, 
\omega)\sim 2\pi\delta(\omega-E_k^{HF})$ as given in 
(\ref{eq:elementary excitations18}), we find (\ref{eq:density fluc2})
 reduces to
\begin{equation}
\chi^0_{nn}(q, \omega)\sim\int d{\bf k}{N^0(E^{HF}_k) 
- N^0(E^{HF}_{k-q})\over (E^{HF}_k - E^{HF}_{k-q})-\omega}.
\label{eq:density fluc3}
\end{equation}
We note that $\chi_{nn}^0$ has a continuum of poles given by $\omega 
= (E_k^{HF} - E^{HF}_{k-q})$.  It is easy to understand the physics 
which gives rise to this ``ideal-gas'' spectrum.
The Fourier transform of the local density operator in 
(\ref{eq:elementary excitations2}) 
is given by (for a uniform system)
\begin{equation}
{\hat n}_q = \sum_k {\hat a}_k^ +{\hat a}_{k-q}\label{eq:density 
fluc4}
\end{equation}
Clearly ${\hat n}_q$ creates a ``particle-hole'' density fluctuation 
with the following features:
\begin{eqnarray}
\mbox{change in energy} & : &  E_k - E_{k-q} = \omega   \nonumber\\
\mbox{change in momentum} & : &  {\bf k} - ({\bf k} - {\bf q}) = {\bf 
q}.\label{eq:density fluc5}\end{eqnarray}
The spectrum (\ref{eq:density fluc3}) describes a broad incoherent 
superposition of particle-hole states \cite{ref:Pinesnoz} and is not 
a true collective mode, such as we discuss next. 

The mean field approximation (MFA) for the density response is also 
called other names: SCF (self-consistent field), RPA (randon phase 
approximation), but all involve the same physics.
The MFA was introduced by Bohm and Pines in the period 1951-1953, 
work which had a pivotal effect in our understanding collective 
effects in all many-particle systems.
We recall the linear response expression in (\ref{eq:elementary 
excitations7}), where $\chi_{nn}$ is the {\it full} density response 
function for interacting Bose gas.
If we introduce the self-consistent Hartree mean-field:
\begin{equation}\delta V_{\mbox{Hartree}} (1) =  \int d1' 
v(1-1')\delta n(1') = g\delta n(1), \label{eq:density 
fluc6}\end{equation}
then we can approximate the linear response equation in 
(\ref{eq:elementary excitations7}) by
\begin{equation}\delta n(1) = \int d1' \chi^0_{nn} (1, 1') [V_{\rm 
ex}(1') +\delta V_{\mbox{Hartree}}(1')].
\label{eq:density fluc7}\end{equation}
The system is assumed to respond as if the atoms propagate 
independently (as described by $\chi^0)$ but are moving in an 
effective field $V_{eff}(1')$.  In a uniform system, the Fourier 
transform of 
(\ref{eq:density fluc7}) is
\begin{equation}\delta n(q, \omega) = \chi^0_{nn} (q, 
\omega)[V_{q\omega} + g\delta n(q, \omega)],\label{eq:density fluc8}
\end{equation}
which gives the well-known MFA expression for $\chi_{nn}$ 
\cite{ref:Fetwal,ref:Pinesnoz}:
\begin{equation}\chi_{ nn}(q, \omega) = {\chi^0_{nn}(q, \omega) \over 
1-g\chi^0_{nn}(q, \omega)}.\label{eq:density fluc9}\end{equation}
This result for  $\chi_{nn}(q, \omega)$ may have new poles given by 
zeros of the denominator, 
\begin{equation}
1 - g\chi^0_{nn}(q, \omega) = 0.\label{eq:density fluc10}
\end{equation}
Assuming that it is distinguisable from the incoherent ideal gas 
density fluctuation spectrum given by 
(\ref{eq:density fluc3}), 
this pole at $\omega = E^{coll}_q$ is called a zero sound mode (a 
plasmon in charged systems).
This language was first introduced in 1957 by Landau in Fermi systems 
\cite{ref:Pinesnoz} but the concept is generally applicable in any 
interacting many body system.
Physically, it is clear that zero sound is a ``collisionless'' 
density oscillation arising from dynamic self-consistent mean fields.

We make a few comments about such ``zero sound'' collective modes:

\begin{enumerate}
\item $E^{coll}_q$ and $E^{sp}_q$  (the poles of $G_1(q, \omega)$) 
are {\it both} states of an interacting many-particle system.
However the collective mode disappears if there are no interactions, 
while the single-particle excitations still exist in a 
non-interacting gas.

\item Because of the low density, dynamic mean fields in normal 
systems are too weak to allow the existence of a well-defined 
(weakly-damped) zero sound mode in the Bose gases of current interest 
(for $T > T_{\rm BEC}$).
If $g\chi^0_{nn}(q, \omega) \ll 1$,  we can then approximate 
$\chi_{nn}(q, \omega) \simeq \chi^0_{nn} (q, \omega)$.  The situation 
is quite different for $T < T_{\rm BEC}$, when a coherent mean-field 
due to the condensate is present.

\item One {\it also} expects a collective pole to appear in  
$\chi_{nn}(\bf q, \omega)$ in the collision-dominated hydrodynamic 
region.
However this sound wave pole is induced by rapid collisions producing 
local equilibrium, as we discuss in Sections~\ref{sec:Hydrodynamic 
oscillations in a trapped Bose gas} - \ref{sec:First and second sound 
in a uniform Bose-condensed gas}.
Ordinary sound is not described by the MFA discussed above, ie, it is 
not the result of dynamic mean fields.
\end{enumerate}

\section{Green's function formulation of excitations in a 
Bose-condensed system}
\label{sec:Green's function formulation}

All the formalism we have been discussing for $T>T_{\rm BEC}$ can be 
extended in a natural way to Bose-condensed systems 
\cite{ref:Bel2,ref:Grif2,ref:Fetwal}
making use of the fundamental decomposition which separates out the 
condensate and non-condensate parts of the quantum fields:
\begin{equation}
{\hat\psi} ({\bf r}) 
= \langle{\hat\psi}({\bf r})\rangle + {\tilde\psi}({\bf r}).
\label{eq:Green function1}
\end{equation}
Here the average is over a restricted ensemble 
\cite{ref:Bog2,ref:Hohmar}
consistent with $\langle{\hat\psi}\rangle \neq 0$. The most profound 
way of doing this is to add a symmetry-breaking field
\begin{equation}
{\hat H}_{sb}(t) = \int d{\bf r} [\eta({\bf r}, t){\hat\psi}^+({\bf 
r}) + \mbox{H.C.}]
\label{eq:Green function2}\end{equation}
and work with ${\hat H}_{tot} = {\hat H}_{system} + {\hat H}_{sb}$,   
taking the limit $\eta\rightarrow 0$ at the end.
This gives the system a ``hunting license'' \cite{ref:Bog2} to have 
finite value of $\langle{\hat \psi}\rangle$  and one finds 
\begin{equation}
\langle{\hat\psi}\rangle_{sb} = 0 \  \mbox{for}\  T > T_{\rm BEC} \ ; 
\ \langle{\hat\psi}\rangle_{sb}\neq 0\  \mbox{for}\  T < T_{\rm BEC}.
\label{eq:Green function3}\end{equation}
A key point is that if $\langle{\hat\psi}\rangle_{sb}\neq 0$, then a 
direct consequence is the existence of ``anomalous'' or 
``off-diagonal'' propagators \cite{ref:Bel2}
\begin{eqnarray}
&\langle&{\tilde\psi}(1){\tilde\psi}(1')\rangle_{sb} \neq 0 
\nonumber\\
&\langle&{\tilde\psi}^+(1){\tilde\psi}^+(1')\rangle_{sb} \neq 0 
.\label{eq:Green function4}
\end{eqnarray}
These describe new condensate-induced correlations between 
non-condensate atoms at different space-time points $1$ and $1'$.
In a sense, these anomalous correlation functions are as important as 
the macroscopic wavefunction $\Phi(1)$.  We also note that ${\tilde 
m}(1)\equiv \langle {\tilde\psi}(1){\tilde\psi}(1)\rangle_{sb}$ is 
the ``pair'' function that Burnett~\cite{ref:Bur} {discusses in 
detail, using a different formalism.
>From now on, we leave the symmetry-breaking label on the averages 
implicit.

In addition, if $\langle{\hat\psi}\rangle \neq 0, $ we find that 
correlation functions involving three non-condensate field operators 
can be finite
\begin{equation}
\langle{\tilde\psi}(1){\tilde\psi}(2){\tilde\psi}(3)\rangle \neq 
0.\label{eq:Green function5}
\end{equation}
In particular, $\langle {\tilde n}(1){\tilde\psi}(1')\rangle \neq 
0$.  This describes the condensate-induced coupling of non-condensate 
density fluctuations ${\tilde n} = {\tilde\psi}^+{\tilde\psi}$
and the single-particle field fluctuations (see Section~\ref{sec:Why 
are excitations in a Bose fluid so interesting}).

Clearly one has to work with a $2\times 2$ matrix single-particle 
propagator $G_1$ when $\langle{\hat\psi}\rangle \neq 0$, namely
\begin{eqnarray}
G_{\alpha\beta} 
&=& \left(
\begin{array}{ll}
\langle \hat\psi(1) \hat\psi^+ (1') \rangle & \langle \hat\psi(1) 
\hat\psi(1') \rangle \\
\langle \hat\psi^+(1) \hat\psi^+(1') \rangle & \langle \hat\psi^+(1) 
\hat\psi(1')\rangle 
\end{array}
\right) \nonumber \\
&=& \left( 
\begin{array}{ll} 
\Phi(1) \Phi^*(1') + \tilde G_{11} & \Phi(1) \Phi(1') + \tilde G_{12} 
\\
\Phi^*(1) \Phi^*(1') + {\tilde G}_{21} & \Phi^*(1) \Phi(1') + {\tilde 
G}_{22}
\end{array}
\right), \label{eq:Green function6}
\end{eqnarray}
where we have introduced a  $2\times 2$ matrix Green's function for 
the non-condensate atoms
\begin{equation}
{\tilde G}_{\alpha\beta} \equiv \left( \begin{array}{cc}{\tilde 
G}_{11} & {\tilde G}_{12}\\ 
{\tilde G}_{21} & {\tilde G}_{22} \end{array}\right) .\label{eq:Green 
function7}
\end{equation}
Beliaev \cite{ref:Bel2} effectively showed (in modern matrix notation)
\begin{equation}
{\tilde G}_{\alpha\beta} = G_0\delta_{\alpha\beta} + G_0 
\Sigma_{\alpha\delta}{\tilde G}_{\delta\beta},\label{eq:Green 
function8}
\end{equation}
where we use the standard convention that repeated indices $(\delta = 
1, 2)$ are summed over.
This is the famous Dyson-Beliaev equation, involving a 2 x 2 matrix 
self-energy $\Sigma_{\alpha\delta}$.
Clearly all components ${\tilde G}_{\alpha\beta}$ will share the same 
single-particle excitation spectrum.
In a 
uniform system, we have 
\begin{equation}
{\tilde G}_{\alpha\beta}(q,\omega) = 
G_0(q,\omega)\delta_{\alpha\beta}+ G_0( 
q,\omega)\Sigma_{\alpha\delta}(q, \omega){\tilde 
G}_{\delta\beta}(q,\omega),\label{eq:Green function9}
\end{equation}
which is a set of linear algebraic equations which are easy to solve 
for ${\tilde G}_{11}(\sim{\tilde G}_{22})$ and ${\tilde G}_{12} 
(\sim{\tilde G}_{21})$.

We will now use this Beliaev formalism to discuss various simple 
approximations for the self-energies $\Sigma_{\alpha\beta}$.
The interaction energy in (\ref{eq:elementary excitations3}), namely
\begin{equation}
V_{int} = {1\over 2}g \int d{\bf r}{\hat\psi}^+({\bf 
r}){\hat\psi}^+({\bf r}){\hat\psi}({\bf r}){\hat\psi}({\bf 
r}),\label{eq:Green function10}
\end{equation}
splits into various distinct contributions when we use 
(\ref{eq:Green function1}) 
to separate out the condensate parts (see Fig.
2).
At $T = 0, $ we can ignore the $V_3$ and $V_4$ contributions because 
so few atoms are in non-condensate.
This is the famous Bogoliubov approximation \cite{ref:b}, and it is 
equivalent to the  linearized GP theory.
In Beliaev language, the Bogoliubov self-energies are shown in Fig. 3.
For a uniform gas, these give \cite{ref:Bel2,ref:Fetwal}
\begin{eqnarray} 
{\tilde G}_{11}(p, \omega) &= &{\omega +\epsilon_p + n_cg\over 
\omega^2 - \left[\epsilon^2_p +2\epsilon_p n_cg\right]} \nonumber\\
{\tilde G}_{12}(p, \omega) &= &{-n_cg\over \omega^2 - 
\left[\epsilon^2_p +2\epsilon_p n_cg\right]}\ .\label{eq:Green 
function11}
\end{eqnarray}
These Bogoliubov single-particle Green's functions contain the same 
physics as discussed in Fetter's lectures \cite{ref:a}.
They clearly have  poles at the frequencies $\omega = \pm E_p$, where
\begin{equation}
E_p = (\epsilon^2_p +2\epsilon_p n_cg)^{1/2}.\label{eq:Green 
function12}
\end{equation}
At low $p$, the single-particle excitation is phonon-like $E_p = 
c_{Bog}p$, with the Bogoliubov phonon velocity
\begin{equation} 
c^2_{Bog} = {n_c g\over m}.
\label{eq:Green function13}
\end{equation}
It turns out that this simple $T=0$ Bogoliubov theory already 
exhibits most of the structure which will always arise in 
Bose-condensed systems.
This is why it plays the role of the ``H-Atom'' in discussions of 
Bose-condensed fluids \cite{ref:Grif2}.

In this approximation,  as we have stressed, the starting point 
assumes that all atoms are in the condensate and hence $n_c = n$.  
However, we can still use ${\tilde G}_{11} ({\bf p}, \omega)$ in 
(\ref{eq:Green function11}) to estimate the depletion.
This makes use of the relation 
\begin{eqnarray}
{\tilde n} &=& n-n_c \sim{\tilde G}_{11} (1, 1) \nonumber\\
&=& \int d{\bf q}\int d\omega N^0 (\omega-\mu){\tilde A}_{11}({\bf 
q}, \omega),\label{eq:Green function14}
\end{eqnarray}
which is an exact formula for ${\tilde n}$ (see Ch.4 of Ref. 
\cite{ref:Grif2}) in terms of the so-called single-particle {\it 
spectral density} [see (\ref{eq:density fluc2})].
The 
spectral density given by the Bogoliubov approximation above is
\begin{equation} 
A_{Bog}({\bf q}, \omega) 
=  u^2_q\delta (\omega-E_q) - v_q^2\delta(\omega + E_q),
\label{eq:Green function15}
\end{equation}
showing the characteristic negative energy pole at $\omega=-E_q.$
This last feature emphasizes that creating an atom in a 
Bose-condensed system involves a coherent weighted combination of
creating an excitation  {\it and} destroying an excitation.
Calculating the depletion, one finds
\begin{eqnarray}
{\tilde n} &=& \int d{\bf q} \int d\omega N^0(\omega)[u_q^2\delta 
(\omega - E_q) - v_q^2\delta(\omega + E_q)] \nonumber\\
&= &\int d{\bf q}\left\{N^0(E_q)u^2_q 
-N^0(-E_q)v^2_q\right\}\nonumber\\
&=& \int d{\bf q}\left\{v^2_q + N^0(E_q)(u^2_q + 
v^2_q)\right\}\nonumber\\
&=& {8\over 3} n_c \left(n_c a^3\over\pi\right)^{1/2},
\label{eq:Green function16}
\end{eqnarray}
a result first obtained by Bogoliubov \cite{ref:b} and reviewed in 
detail by Fetter \cite{ref:a}.
We have made use of the key identity which the Bose distribution 
satisfies, 
\begin{equation} 
N^0(-E) = - [N^0(E)+1].
\label{eq:Green function17}
\end{equation}
We see from 
(\ref{eq:Green function16})
that the $T=0$ depletion of the condensate formally arises from the 
negative energy pole in ${\tilde G}_{11}({\bf q}, \omega)$.  

It is easy to evaluate the density response function 
$\chi^0_{nn} (q, \omega)$ in (\ref{eq:density fluc1})
 using the Bogoliubov spectral density in (\ref{eq:Green function15}).
The negative energy poles are seen to give rise to new poles at 
$\omega = E_k + E_{k-q}$ which have finite weight even at $T=0$.
These are in addition to the ``normal'' particle-hole poles of 
$\chi_{nn}^0$ at $\omega = E_k - E_{k-q}$, as given by 
(\ref{eq:density fluc3}).

\section{Beyond the Bogoliubov approximation: classification in terms 
of self-energies}
\label{sec:Beyond the Bogoliubov approximation}

At finite $T$, a lot of atoms are thermally excited out of the 
condensate in a dilute Bose gas.  These ``normal'' self-energies must 
be included.  In the Beliaev self-energy formalism, this was first 
done by  Popov 
in 1965 \cite{ref:Popov1}.  We have to add the {\it ordinary} 
Hartree-Fock self-energies, as shown in Fig.4.  This gives the 
first-order Popov approximation (the excited atoms are treated as as 
ideal Bose gas, and thus the approximation is not self-consistent)
\begin{equation}
\begin{array}{ll}
\Sigma_{11}^{Popov} (p, \omega) &= 2n_cg+2{\tilde n}^0g \nonumber\\
\Sigma_{12}^{Popov} (p, \omega) &= n_cg.\label{eq:Beyond1}
\end{array}
\end{equation}
These self-energies lead to [compare with (\ref{eq:Green function11})]
\begin{equation} 
{\tilde G}_{11} (p, \omega) = {\omega + \epsilon_p+\Delta\over 
\omega^2 - E_p^2} \ ; \ {\tilde G}_{12} (p, \omega) = {-\Delta\over 
\omega^2 - E^2_p}\ , 
\label{eq:Beyond2}\end{equation}
with
\begin{equation} \Delta \equiv \mu_{\rm HP} - 2{\tilde n}^0g \ ; \ 
E^2_p \equiv \epsilon^2_p + 2\epsilon_p\Delta , 
\end{equation}\label{eq:Beyond3}
where
\begin{equation} \mu_{\rm HP} \equiv \Sigma_{11}(p = 0, \omega = 0) - 
\Sigma_{12} (p = 0, \omega = 0).
\label{eq:Beyond4}
\end{equation}

The ``chemical potential''  $\mu_{\rm HP}$ as defined in 
(\ref{eq:Beyond4}) 
here was  introduced by Hugenholtz and Pines \cite{ref:Hugpines} at 
$T = 0$ (and generalized to $T\neq 0$ later \cite{ref:Hohmar}).
HP showed that if ${\tilde G}_{\alpha\beta}(p, \omega)$ had a gapless 
excitation spectrum in the $p \rightarrow 0, \omega\rightarrow 0$ 
limit, then the true chemical potential $\mu$ must satisfy $\mu = 
\mu_{\rm HP}$.
In the Popov approximation, calculation gives \cite{ref:Grif3}
\begin{equation}
\mu_{\rm HP} = 2n_cg + 2{\tilde n}^0g - n_cg = n_cg +  2{\tilde n}^0 
g,\label{eq:Beyond5}
\end{equation}
and thus $\Delta = n_cg$, where $n_c(T)$ is the ideal gas condensate 
density at temperature $T$.
The generalized GP equation determines the chemical potential, which 
we find convenient to denote by $\mu_c$.  In the Popov approximation, 
this equation is given by 
\begin{equation}
\left\{-{\nabla^2\over 2m} + U_{\rm ext} + gn^0_c + 2g{\tilde n}^0 
\right\}\Phi =\mu_c\Phi, \label{eq:Beyond6}\end{equation}
and hence in a uniform gas, we see that $\mu_c = \mu_{\rm HP}$.
In the  Popov approximation, the single-particle Green's functions  
${\tilde G}_{\alpha, \beta}$ 
in (\ref{eq:Beyond2}) have poles given by
\begin{equation}
E_p = [\epsilon^2_p +2\epsilon_p n_c(T)g]^{1/2}.\label{eq:Beyond7}
\end{equation}
This is formally the same as the $T=0$ Bogoliubov excitation 
frequency, except that now  $n_c(T)$ is temperature-dependent.
For a more detailed discussion, see Chapter 3 of the recent review by 
Shi and Griffin \cite{ref:Shigrif}.

The {\it full} Hartree-Fock Bogoliubov (HFB) approximation involves 
calculating  $\Sigma_{\alpha\beta}$ self-consistently using the 
complete matrix $2\times 2$ Beliaev propagator, as shown in Fig. 5.  
We note that above $T_{\rm BEC}$, the HFB reduces to the usual 
self-consistent Hartree-Fock approximation.
Moreover,  the HFB is the best {\it single-particle} approximation 
for a Bose-condensed system, in a variational sense.
Within this limitation, the HFB will give the best single-particle 
approximation for the thermodynamic properties.

However, there is one ``bad" aspect of the full HFB which was noticed 
in the 1960's and which has been the subject of a several recent 
theoretical papers on Bose-condensed gases.
Namely, the HFB does not obey the Hugenholtz-Pines (HP) theorem.
In a uniform gas, this means that the single-particle excitations 
will have an energy gap in the long wavelength limit {\it{q}} 
$\rightarrow$ 0.
It is easy to check this by calculating the two chemical potentials 
introduced earlier:
\begin{eqnarray}\mu_{\rm HP} &=& 2 (n{_c}+ {\tilde n}) g - 
(n{_c}+{\tilde m})g\nonumber\\
&=& g (n+{\tilde n}- {\tilde m}), \mbox{using}\ n = n{_c}+ {\tilde 
n}.\label{eq:Beyond8}\end{eqnarray}
The static equation of motion for $\Phi = \sqrt{n_c}$ using the HFB 
approximation gives
\begin{equation} \left[-{\nabla ^2\over 2m}+U_{\rm 
ext}+gn_{c}+2g\tilde{n}+g\tilde{m}\right]{\sqrt n_{c}}= \mu_{c}\sqrt 
n_{c}\rightarrow\mu_{c} = 
g(n+\tilde{n}+\tilde{m}).\label{eq:Beyond9}\end{equation}
One ``solution'' of this problem is to leave out $\tilde{m}$ (which 
is the source of the problem) but keep $\tilde{n}$.
This corresponds to the self-consistent HFP(opov) approximation 
\cite{ref:Grif3,ref:Hutchzargrif,ref:Dalgiopitstr}
which, while approximate,  has several nice features.
It gives the correct excitation spectrum for both $T \rightarrow$ 0 
and $T > T_{\rm BEC}$.
Moreover, the spectrum is gapless at all temperatures.

The origin of this ``problem" with the HFB is clear.
The HFB keeps all self-energies which are first order in the 
interaction $g$.
However, by computing these self-energy diagrams using 
self-consistent ${\tilde G}_{\alpha\beta}$ propagators, one is 
clearly bringing in terms to all orders in $g$.
This brings one into dangerous territory! Moreover, one easily can 
check that $\tilde{m}$ must be at least of order $g$.
Thus the $g\tilde{m}$ contribution to $\Sigma _{12}$ is at least of  
$O(g^2$).
This suggests that to ``fix-up'' the HFB as a theory of excitations, 
we have to include all self-energy contributions to at least second 
order in $g$.

Moreover, it is not obvious that it is consistent to calculate 
$\tilde {n}$ self-consistently \cite{ref:Hutchzargrif} and ignore 
$\tilde{m}$, since the lowest order interaction contributions to both 
quantities are found to be of the same order.  Indeed, one can show 
for a uniform gas that keeping the lowest order asymptolic  
correction to the ideal gas result gives $\tilde{n}=\tilde{n}_{\rm 
cr}-\tilde{m}$, where $\tilde{n}_{\rm cr} = n (T/T_{\rm 
BEC})^\frac{3}{2}$
is the critical density of an ideal gas at temperature $T$ 
\cite{ref:Fedshl, ref:Shigrif}.  However, we note that an expansion 
around the ideal gas results is not necessarily a good approximation 
in trapped gases (see also the discussion of the results in 
(\ref{eq:two-fluid7}).  At $T = 0$, the Popov approximation is a poor 
guide for corrections to the Bogoliubov approximation since a 
calculation analogous to (\ref{eq:Green function16}) gives  
$\tilde{m}=3\tilde{n}$ \cite{ref:Shigrif}.

So, finally,  we come to the work of Beliaev \cite{ref:Bel1} who, in 
1957, evaluated the $\Sigma_{\alpha\beta}$ self-energies keeping all 
contributions up to order $g^2$.
This was originally done at $T = 0$.  For finite  temperatures $T\sim 
T_{\rm BEC}$, Beliaev's work has been recently extended by several 
authors 
\cite{ref:Popov1,ref:Shigrif,ref:Fedshl}.
When one includes the second-order diagrams for 
$\Sigma_{\alpha\beta}$, we note that one must be careful to  also 
treat the HFB first-order diagrams correctly to order $g^2$ since 
there are many terms which cancel.

As expected, the  Beliaev-type second order calculation cures all the 
problems of the first-order HFB approximation.
We make a few comments on the results for a uniform Bose gas.
The self-energies are found to be

\begin{eqnarray}\Sigma_{11}(q,\omega) & =& 2n{_c} g{_R} + 2{\tilde 
n}^{(1)}g + A(q, \omega)g^2\nonumber\\
\Sigma_{12} (q,\omega)&=& n_cg_R + {\tilde m}^{(1)}g + B(q, 
\omega)g^2,\label{eq:Beyond10} \end{eqnarray}
with the renormalized interaction defined by
\begin{equation} g{_R}= g \left[1+g\int {d{\bf k}\over 
(2\pi){^3}}{1\over 
2\epsilon_{k}}\right]\label{eq:Beyond11}\end{equation}
and
\begin{equation}\tilde{m}^{(1)} = -n{_c}g \int {d{\bf k}\over 
(2\pi){^3}}\left[{2N^{0}(E_k)+1\over 
2E_{k}}\right].\label{eq:Beyond12}\end{equation}
Here the anomalous and normal densities, 
$\tilde{m}^{(1)}=\tilde{G}_{12}(1,1)$ and 
$\tilde{n}^{(1)}=\tilde{G}_{11}(1,1)$, are computed using the HFP 
approximation for $\tilde{G}_{\alpha\beta}$.
$g_R$ is the second-order approximation to the $t$-matrix, as shown 
in Fig.~6.
Using these results to calculate the chemical potential, we find (for 
$q, \omega\rightarrow 0)$
\begin{eqnarray}
\mu_{\rm 
HP}&=&2n_{c}g_{R}+2\tilde{n}^{(1)}g-n_{c}g_{R}-\tilde{m}^{(1)}g
+(A-B)g^{2}\nonumber\\
&=&n_{c}g_{R}+2\tilde{n}^{(1)}g+\tilde{m}^{(1)}g.
\label{eq:Beyond13}
\end{eqnarray}
This last result gives precisely the HFB result for $\mu_{c}$, and 
hence the HP relation is satisfied.
In the last line of (\ref{eq:Beyond13}), we have used the key result 
that for $q, \omega\rightarrow 0$, $(A-B)g^{2}=2\tilde{m}^{(1)}g$, as 
proven at all temperatures by Talbot and Griffin \cite{ref:TalGrif}.

We note that one can rewrite the expression in 
(\ref{eq:Beyond13}) 
in the alternative form
\begin{equation}\mu = g\left[n_c + {\tilde n}^{(1)} + {\tilde 
m}_R^{(1)}\right],\label{eq:Beyond14} \end{equation}
where we have introduced a ``renormalized'' anomalous density
\begin{eqnarray} 
{\tilde m}_R^{(1)} &\equiv& {\tilde m}^{(1)} - n_cg \int{d{\bf 
k}\over (2\pi){^3}}{1\over 2\epsilon_{k}}\nonumber\\
&=& - n_cg\int {d{\bf k}\over (2\pi){^3}}\left[{2N^{0}(E_k)+1\over 
2E_{k}}- {1\over 2\epsilon_{k}}\right].\label{eq:Beyond15}
\end{eqnarray}
In contrast to ${\tilde m}^{(1)}$ in (\ref{eq:Beyond12}), ${\tilde 
m}^{(1)}_R$ has no ultraviolet divergence from large $k$ 
contributions.
Proukakis, Burnett and coworkers \cite{ref:Bur} have pointed out that 
the self-consistent ladder diagram approximation for the $t-$matrix 
(see Ref.\cite{ref:Bijstoof} and Ch.4 of Ref.\cite{ref:Shigrif}) can 
be expressed in terms of ${\tilde m}_R$.  In a uniform gas, this 
relation is simply $t = g(1 + {\tilde m}_R/n_c)$.  This suggests that 
the effect of ${\tilde m}$ can be partially included by working in 
terms of the self-consistent $t-$matrix \cite{ref:Bur}.

\section{Physics of the Beliaev approximation}
\label{sec:Physics of the Beliaev approximation}

We have seen in 
Section~\ref{sec:Beyond the Bogoliubov approximation}
that Beliaev's second-order calculation satisfies the HP theorem.
It has another feature whose significance was only realized in the 
early 1960's.
Using the above $T=0$ Beliaev expression (\ref{eq:Beyond14}) for the 
chemical potential $\mu$, one finds at $T=0$ (a result first found by 
Lee and Yang using a different approach)
\begin{equation}
\mu_{\rm Bel} = ng \left[1+{32\over 
3}\left({na^3\over\pi}\right)^{1/2}\right], 
\label{eq:approximation1}
\end{equation}
where we have used [see (\ref{eq:Green function16})]
\begin{equation} n=n_c+n_c{8\over 
3}\left({n_ca^3\over\pi}\right)^{1/2}\equiv n_c +{\tilde n}.
\label{eq:approximation2}\end{equation}
One also finds that the phonon pole of ${\tilde 
G}^{Bel}_{\alpha\beta}$ has a velocity given by
\begin{equation} 
c^2_{\rm Bel} 
= {4\pi na\over 
m^2}\left[1+16\left({na^3\over\pi}\right)^{1/2}\right], 
\label{eq:approximation3}
\end{equation}
again expressed in terms of the total density $n$.
One may easily check that
\begin{equation} 
c^2_{\rm Bel} = {n\over m}{d\mu(n)\over dn}, \label{eq:approximation4}
\end{equation}
that is, the Beliaev phonon excitation at $T=0$ has a velocity 
precisely equal to the {\it compressional} sound velocity.

This last result raises the basic question: why does the elementary 
excitation (pole of ${\tilde G}_{\alpha\beta})$ have a velocity 
corresponding to that of a density fluctuation (pole of 
$\chi_{nn})$?  Later this strange identity was proven to be correct 
at $T=0$ to {\it all} orders of perturbation theory by Gavoret and 
Nozi${\grave{\rm e}}$res \cite{ref:Favnoz}.
To be precise, at $T=0, $ in any Bose-condensed fluid (liquid or 
gas), GN showed (for $q, \omega\rightarrow 0)$
\begin{equation}{\tilde G}_{\alpha\beta}(q, 
\omega)\sim{a\over\omega^2-c^2q^2} \ ; \ \chi_{nn}(q, \omega) 
\sim{b\over\omega^2-c^2q^2}\ ^, 
\label{eq:approximation5}\end{equation}
where $c^2$ is given by the compressibility as in 
(\ref{eq:approximation4}).
We discuss this equivalence in Section~\ref{sec:Why are excitations 
in a Bose fluid so interesting}.

Let us consider the full time-dependent HFB equation of motion for 
$\Phi ({\bf{r}},t)\equiv \langle{\hat\psi} ({\bf{r}})\rangle_{t}$, 
given by
\begin{equation} 
i {\partial\Phi({\bf r},t)\over \partial t}
=\left[-{\nabla^2\over2m}+U_{\rm ext}({\bf r})+gn_{c}({\bf r}, t)
+ 2g\tilde{n}({\bf r}, t)+g\tilde{m}({\bf r,} t)\right]
\Phi({\bf r},t). 
\label{eq:approximation6}
\end{equation}
There are various approximations to this key equation which have been 
used in recent discussions of trapped Bose gases:

\begin{itemize}
\item {\it GP:} ignore $\tilde{n}$ and $\tilde{m}$ completely.
This is valid at $T = 0$ 
\cite{ref:Dalgiopitstr,ref:Edrupburdodclark}.
\item {\it Static HFP:}  Keep $n_{c}({ \bf r},t)$ but set 
$\tilde{n}({\bf r},t) =\tilde{n}_{0}({\bf r})$ and $\tilde{m}({\bf 
r}, t)=0$ 
\cite{ref:Hutchzargrif,ref:Dodburedclark}.
\item {\it Static HFB:} Keep $n_{c}({\bf r},t)$ but set 
$\tilde{n}({\bf r},t)=\tilde{n}_{0}({\bf r})$ and 
$\tilde{m}({\bf{r}},t)=\tilde{m}_{0}({\bf r})$ \cite{ref:Hohmar, 
ref:Grif3}.
As we have noted, this produces an energy gap in the excitation 
spectrum ( since $\mu_{\rm HP}\neq\mu_{c})$.
\item {\it Dynamic HFB:} Treat all dynamic mean-fields due to 
$n_{c}({\bf r}, t),\tilde{n}({\bf r,} t)$ and $\tilde{m}({\bf r}, t)$ 
on an equal basis in a generalized mean-field calculation of the 
density response function $\chi_{nn}  (1,1^{\prime})$.
One finds \cite{ref:Chunggrif} that the poles of $\chi_{nn}$ given by 
this kind of calculation $(\chi_{nn}\sim{1\over \omega^{2}-E^{2}})$ 
are {\it identical} to the poles of $\tilde{G}_{\alpha\beta}$ given 
by the second-order Beliaev approximation 
$(\tilde{G}_{\alpha\beta}\sim {1\over \omega^{2}-E^{2}}$).
\end{itemize}

Recent work on excitations at $T\neq$0 in trapped gases is easily 
understood in terms of such a generalized linear response calculation 
of 
$\chi_{nn}$.  
For example, Minguzzi and Tosi \cite{ref:Mintosi} only keep 
fluctuations 
in $\delta n_{c}$ and $\delta\tilde{n}$ but $\delta\tilde{m}$ 
is completely ignored (dynamic HFPopov); 
in contrast, Giorgini \cite{ref:Giorgini} keeps fluctuations in both 
$\delta\tilde{n}$ and $\delta\tilde{m}$ but only those induced by 
the condensate fluctuations $\delta n_{c}$.

Let us briefly discuss the excitation frequencies in trapped gases at 
finite 
temperatures~\cite{ref:Bur,ref:Ketterle}.
For $T\leq 0.4$ $T_{\rm BEC}$, GP theory $(T = 0)$ predictions 
\cite{ref:Edrupburdodclark} are in excellent agreement with the JILA 
\cite{ref:Jimensmatwiecor} and MIT 
\cite{ref:Mewandvankurndurftownket} experiments.
For $T\geq 0.6$ $T_{\rm BEC}$, however, there are significant 
differences between most current theoretical results and the measured 
excitation energies  ~\cite{ref:Hutchzargrif,ref:Dodburedclark}.
The problem clearly is that mean-field theories must include the 
collective dynamics of {\it both} the condensate and the 
non-condensate.
The Beliaev treatment does this,  as do the papers 
\cite{ref:Mintosi,ref:Giorgini}
mentioned above, as well as work based on a collisionless kinetic 
equation 
\cite{ref:Zargrinik,ref:Bijlsmastoof}.
These improved treatments can correctly obtain the in-phase rigid 
oscillation of the condensate and non-condensate equilibrium profiles 
(the so-called Kohn mode), as discussed in 
Section~\ref{sec:In-phase oscillation of the condensate and 
non-condensate}.

\section{Why are excitations in a Bose fluid so interesting?}
\label{sec:Why are excitations in a Bose fluid so interesting}

We have seen in 
Section~\ref{sec:Physics of the Beliaev approximation}
that at $T = 0$, the elementary excitations have the same spectrum as 
density fluctuations.
In fact, one can prove this equivalence to be a direct consequence of 
the Bose broken symmetry, $\Phi({\bf{r}})\equiv 
\langle{\hat\psi}\rangle\neq 0$.
Thus, this equivalence is valid at all $T \leq T_{\rm BEC}$.
Of course, as soon as $T \neq 0$, the phonon velocity is no longer 
simply given by the compressional sound velocity 
(\ref{eq:approximation4}) as found at $T = 0$.
This fundamental equivalence was only proven and understood in the 
early 1960's (Gavoret and Nozi${\grave{\rm e}}$res \cite{ref:Favnoz}, 
Hohenberg and Martin \cite{ref:Hohmar}, Bogoliubov \cite{ref:Bog2}).
It was implicitly assumed  by Landau \cite{ref:Land} in his 1941 
paper and later by Feynman 
\cite{ref:Feynman1}
in his 1953--54 papers on excitations.
Of course, the clear distinction between elementary excitations and 
density fluctuations in many-body systems was only clarified in the 
period starting from 1957.
Only then were Bose fluids realized to be very special 
\cite{ref:Bog2,ref:Hohmar,ref:Favnoz}.
Today, we understand this key hidden assumption of the work of Landau 
and of Feynman has its microscopic basis in $\Phi({\bf r})\neq 0$.

Let us try to understand the origin of this equivalence of the field 
fluctuation and density fluctuation spectrum.
The density operator in 
a Bose-condensed system can be decomposed 
using~(\ref{eq:Green function1})
as follows:
\begin{eqnarray}
\hat{n}({\bf r})&=&\hat{\psi}^{+}({\bf r})\hat{\psi}({\bf r}) \cr
&=&|\Phi_{0}({\bf  r})|{^2}+\Phi_{0}({\bf r}){\tilde\psi}^{+}({\bf r})
+\Phi^*_{0}({\bf r}){\tilde\psi}({\bf r})
+{\tilde\psi}^{+}({\bf r}){\tilde\psi}({\bf r}),
\label{eq:excitations1}
\end{eqnarray}
where the first three terms correspond to $ \hat{n}_{c}({\bf 
r})=n_{c0}({\bf r})+\delta n_{c}({\bf r})$.
Clearly density fluctuations in a Bose-condensed system (ie, 
$\Phi_{0}\neq 0$) have a direct coupling to the single-particle field 
fluctuations, due to the possibility of atoms coming in or out of 
condensate reservoir.
In momentum space, we recall that 
(\ref{eq:excitations1}) 
is equivalent to
\begin{equation} {\hat n}_q= \sum_k{\hat a}_k^ +  {\hat 
a}_{k-q}=|{\hat a}_{0}|^2+({\hat a}_0^+
{\hat a}_{-q}+{\hat a}_{0}{\hat a}^{+}_q)+\tilde{n}_q 
.\label{eq:excitations2}
\end{equation}
Making the same decomposition as in (\ref{eq:excitations1})
 for the density response function (\ref{eq:elementary excitations5})
gives 
\begin{eqnarray}
\chi_{nn}(1,1^{\prime})
&=&\langle \hat{n}(1)\hat{n}(1^{\prime})\rangle \nonumber\\
&=&\langle \hat{n}(1)\rangle \langle \hat{n}(1^{\prime})\rangle 
+\langle \delta n_{c}(1)\delta n_{c}(1^{\prime})\rangle \nonumber\\
& +&\langle \delta n_{c}(1)\delta\tilde{n}(1^{\prime})\rangle 
+\langle \delta\tilde{n}(1)\delta n_{c}(1^{\prime})\rangle \nonumber\\
& +&\langle 
\delta\tilde{n}(1)\delta\tilde{n}(1^{\prime})\rangle.
\label{eq:excitations3}
\end{eqnarray}
We note that in the Bogoliubov approximation, we have $\delta 
\tilde{n}=0$ and the equivalence of $\chi _{nn}$ and 
$\tilde{G}_{\alpha\beta}$ spectrum is then trivial (see also 
\cite{ref:a}) 
\begin{equation}\chi_{nn}(q,\omega)\Rightarrow\chi^{B}_{nn}
(q,\omega) \equiv n_{c0}\sum_{\alpha,\beta}{\tilde 
G}^{B}_{\alpha\beta}(q,\omega).
\label{eq:excitations4}\end{equation}

To illustrate how the hybridization of single-particle and density
fluctuation arises, we discuss the time-dependent HFP
approximation on which Ref.~\cite{ref:Mintosi} is based.
For simplicity, we consider the uniform gas case.
In the MFA of 
Section~\ref{sec:Density fluctuation spectrum in the mean-field
approximation}, we
have
\begin{equation}
\delta \tilde n = \chi^0[V_{\rm ext} + 2 g(\delta n_c + \delta \tilde 
n)],
\label{eq:excitations5}\end{equation}
and hence for $V_{\rm ext} \rightarrow 0$,
\begin{equation}
\delta\tilde n = \frac{\chi^0 2 g \delta n_c}{1 - 2g\chi^0} .
\label{eq:excitations6}\end{equation}
Using this result in the HFP equation of 
motion~(\ref{eq:approximation6}), ie, with
$\delta\tilde m = 0$,
we find
\begin{equation}
(2g\delta\tilde n + g\delta n_c) 
= g\left(\frac{1 + 2g\chi^0}{1 - 2g\chi^0}\right)\delta n_c 
\equiv g'\delta n_c .
\label{eq:excitations7}\end{equation}
In this approximation, which only keeps the $\delta\tilde n$
fluctuations induced by the condensate, the equation of motion for
$\delta\Phi({\bf r}, t)$ reduces to the GP equation, but with
$g \rightarrow g'$.
Thus the characteristic poles will be determined by the zeroes of
\begin{equation}
[\omega^2 - (\epsilon^2_p + 2 g' n_c\epsilon_p)]
\propto [\omega^2 - (\epsilon^2_p + 2 g n_c\epsilon_p)]
[1 - 2g\chi^0] - 4g^2 2n_c \epsilon_p \chi^0 .
\label{eq:excitations8}\end{equation}
This is the same denominator (up to a common factor) exhibited by the
response functions obtained by Minguzzi 
and Tosi~\cite{ref:Mintosi}.
This calculation illustrates how the Bogoliubov single-particle
mode and the zero sound density fluctuation are coupled and 
hybridized.

More generally, one can prove that (at all ${\it{T}})$:
\begin{equation}
\chi_{nn}(1,1')
= \sum_{\alpha,\beta} \int d2\int  d3\,
\Lambda_{\alpha}(1,2)\tilde{G}_{\alpha\beta}(2,3)\Lambda_{\beta}(3,1')
+\tilde{\chi}_{nn}(1,1'),  \nonumber
\label{eq:excitations9}\end{equation}
where $\Lambda_{\alpha}(1,1')$ is a Bose broken-symmetry vertex 
function which vanishes if $n_{c0}=0$.  Using  the ``dielectric 
formalism" developed in the early 1970's (for a review, see Ch.5 of 
Ref. \cite{ref:Grif2}), one can prove that the self-energies 
$\Sigma_{\alpha\beta}$ are such that $\chi_{nn}$ and 
$\tilde{G}_{\alpha\beta}$ have the same poles---and that there are 
none specifically associated with $\tilde{\chi}_{nn}$.
The advantage of this diagrammatic formalism 
\cite{ref:Mawoo,ref:Szepfalusy,ref:Wonggould,ref:Grifcheun,ref:Szepkond}

is that one manifestly sees that the poles of 
$\tilde{G}_{\alpha\beta}$ and $\chi_{nn}$ are identical, within a 
given approximation for the ``building blocks" of the dielectric  
formalism.
We now see why we can study elementary excitations from the density 
fluctuation spectrum when $\Phi_{0}\neq 0$.
To directly measure $\tilde{G}_{1}(1,1')$, we need an external 
perturbation which can change the number of particles in the system.
BEC indirectly provides us with such a probe!

This equivalence of the single-particle excitations
with density fluctuations lies at the heart of the phenomenon of
superfluidity.
It essentially {\it restricts} the possible excited states to density 
fluctuations and thus
ensures the stability of superfluid motion.  However, 
as Nozi\`eres says in a very lucid review article~\cite{ref:Nozieres 
1966},
``that the $T = 0$ superfluid equations merge with those of 
ordinary hydrodynamics does not alter the fact that real
understanding must be based on a microscopic description based on
long-range order induced by a condensate''.

\section{Hydrodynamic oscillations in a trapped Bose gas}
\label{sec:Hydrodynamic oscillations in a trapped Bose gas}

In the previous sections, we have been discussing collective 
oscillations of the condensate and non-condensate in the mean-field 
or {\it collisionless} region.
We now turn to the collision-dominated {\it hydrodynamic} region 
$l\ll\lambda$, where $\lambda$ is the wavelength of the collective 
mode and $l$ is the collisonal mean-free-path of the elementary 
excitations \cite{ref:Pinesnoz}. Experiments in trapped Bose gases 
are just now starting to probe the hydrodynamic two-fluid region but 
this region is very rich in new physics and deserves careful study.

It is unfortunate that the collisionless region at $T=0$ described by 
the Gross-Pitaevskii equation is also commonly referred to as 
``hydrodynamic'' theory 
\cite{ref:Dalgiopitstr,ref:Pethsmi}. I guess we are stuck with this 
terminology, although the terms ``quantum hydrodynamics'' or 
``superfluid hydrodynamics'' would be useful compromises.

We will limit our analysis to {\it finite} temperatures $T\sim T_{\rm 
BEC}$, where the number of atoms in the non-condensate is comparable 
to the condensate $({\tilde N}\sim N_c)$.
We will concentrate on the {\it derivation} of two-fluid hydrodynamic 
equations starting from a  simple microscopic model and try to 
clarify the physics.
These two-fluid hydrodynamic equations are expressed in terms of 
fluctuations of the two components: $n_c({\bf r}, t), \ {\bf 
v}_s({\bf r}, t)$ and
${\tilde n}({\bf r}, t), \ {\bf v}_n({\bf r}, t)$.  We can use them 
to ``derive'' the two-fluid equations in the conventional Landau 
form, which are written in terms of the fluctuations of local 
thermodynamic variables (temperature, entropy, pressure, etc).
We mainly consider the linearized version of the ZGN-type two-fluid 
equations, which give the hydrodynamic normal modes of the coupled 
system of two components.
As a concrete example, we discuss the {\it in-phase} rigid motion of 
the equilibrium density profiles of the condensate and non-condensate.

We need to find equations of motion for both the condensate and the 
non-condensate.  As a first approximation \cite{ref:Zargrinik}, we 
can use the 
time-dependent HF Popov equation of motion  (see 
Section~\ref{sec:Physics of the Beliaev approximation}), for the 
order parameter $\Phi({\bf r}, t):$
\begin{equation} 
i {\partial\Phi({\bf r}, t)\over\partial t} = \left[-{\nabla^2\over 
2m} + U_{\rm ext}({\bf r}) + gn_c({\bf r}, t) + 2g{\tilde n}({\bf r}, 
t)\right]\Phi({\bf r}, t),
\label{eq:HF Popov equation motion1}
\end{equation}
where $n_c ({\bf r}, t) = |\Phi({\bf r}, t)|^2$ and  ${\tilde n}({\bf 
r}, t)$ is the local non-condensate density.
It is clear that this equation of motion for $n_c({\bf r}, t)$ is not 
closed since it involves
\begin{equation} {\tilde n}({\bf r}, t) = {\tilde n}_0({\bf r}) + 
\delta n({\bf r}, t).
\label{eq:HF Popov equation motion2}\end{equation}
If we treat the non-condensate in 
(\ref{eq:HF Popov equation motion1}) 
{\it statically}, this means that the condensate moves in the {\it 
static} Hartree-Fock mean field of non-condensate given by $2g{\tilde 
n}_0({\bf r})$ \cite{ref:Hutchzargrif}
More generally, we need to have an equation of motion for the 
fluctuations in $\delta{\tilde n}({\bf r}, t)$ of the non-condensate 
(ie, the density of the excited atoms), which will be discussed in 
Section~\ref{sec:Dynamics of the non-condensate atoms}.

We can rewrite equation of motion in (\ref{eq:HF Popov equation 
motion1}) in the {\it quantum hydrodynamic} variables for phase and 
amplitude, defined by
\begin{equation} \Phi({\bf r}, t) \equiv \sqrt{n_c({\bf r}, t)} 
e^{i\theta({\bf r}, t)},
\label{eq:HF Popov equation motion3}\end{equation}
where the superfluid velocity is given by 
\begin{equation} {\bf v}_s ({\bf r}, t) 
\equiv{\mbox{\boldmath$\nabla$}\theta ({\bf r}, t)
\over m}.\label{eq:HF Popov equation motion4}
\end{equation}
Using (\ref{eq:HF Popov equation motion3}), (\ref{eq:HF Popov 
equation motion1}) can be shown to be equivalent to (all quantities 
depend on ${\bf r}$ and $t$)
\begin{eqnarray}
{\partial n_c\over\partial t} + \mbox{\boldmath$\nabla$}\cdot(n_c{\bf 
v}_s) &=&0 \nonumber\\
m\left[{\partial {\bf v}_s\over\partial t}+({\bf 
v}_s\cdot\mbox{\boldmath$\nabla$}){\bf v}_s\right] &=& 
-\mbox{\boldmath$\nabla$}\mu_c,\label{eq:HF Popov equation motion5}
\end{eqnarray}
where the {\it superfluid chemical potential} is given by (in the 
dynamic HFP approximation) 
\begin{equation}
\mu_c({\bf r}, t) \equiv -{\nabla^2\sqrt{n_c}\over 2m\sqrt{n_c}} + 
U_{\rm ext}({\bf r}) + gn_c({\bf r}, t)+ 2g{\tilde n}({\bf r}, 
t).\label{eq:HF Popov equation motion6}
\end{equation}	
The first term in (\ref{eq:HF Popov equation motion6}) is the 
``quantum pressure'' term, which is {\it ignored} in the so-called 
Thomas-Fermi (TF) approximation to the condensate equation of motion 
\cite{ref:Dalgiopitstr,ref:Pethsmi}.
If we omit the non-condensate term, equations (\ref{eq:HF Popov 
equation motion5}) and (\ref{eq:HF Popov equation motion6}) describe 
the well-known Gross-Pitaevskii approximation 
\cite{ref:Dalgiopitstr}}.

\section{Dynamics of the non-condensate atoms}
\label{sec:Dynamics of the non-condensate atoms}

Following Ref.\cite{ref:Zargrinik}, we base our discussion on the 
simplest possible {\it kinetic equation} for the distribution 
function $f({\bf r, p},t)$ valid in the semi-classical limit:
\begin{eqnarray} 
k_BT &\gg& 
\hbar\omega_0\ \ (\omega_0 \equiv \mbox{trap frequency}) \nonumber\\
k_BT &\gg& gn.
\label{eq:semi-classical limit1}
\end{eqnarray}
This distribution function is determined by the quantum Boltzmann 
equation
\begin{equation}
\left[{\partial\over\partial t}+{{\bf p}\over 
m}\cdot\mbox{\boldmath$\nabla$}_r - \mbox{\boldmath$\nabla$}U({\bf 
r}, t)\cdot\mbox{\boldmath$\nabla$}_p\right]
f({\bf r, p}, t) = \left.{\partial f({\bf r, p}, t)\over\partial 
t}\right|_{\mbox{collision}},
\label{eq:semi-classical limit2}\end{equation}
where the effective potential
\begin{equation} U({\bf r}, t) = U_{\rm ext}({\bf r}) + 2g[n_c({\bf 
r}, t) + {\tilde n}({\bf r}, t)].\label{eq:semi-classical limit3}
\end{equation}
includes the dynamic HF field.  
Once we solve (\ref{eq:semi-classical limit2}) for $f({\bf r, p}, t), 
$ we can find the {\it non-condensate} density from 
\begin{equation} {\tilde n}({\bf r,} t) = \int{d{\bf p}\over 
(2\pi)^3} f({\bf r, p}, t).
\label{eq:semi-classical limit4}\end{equation}
In our simple model, the excited atoms are still particle-like, ie,
\begin{equation} E_p({\bf r}, t) = {p^2\over 2m} + U({\bf r}, t).
\label{eq:semi-classical limit5}\end{equation}
This last fact leads to many simplifications but restricts the 
results to finite temperatures 
close to $T_{\rm BEC}$.

We now give the explicit form for the collision term in 
(\ref{eq:semi-classical limit2}): 
\begin{equation}
\left.{\partial f\over\partial t}\right|_{\mbox{collision}}= 
C_{22}[f] + C_{12}[f].
\label{eq:semi-classical limit6}\end{equation}
The term $C_{22}[f]$ describes collisions between excited atoms and 
is given by \cite{ref:Uehlinguhlenbeck},
\begin{eqnarray}
C_{22}[f] &=& 2g^2 \int{d{\bf p}_1\over(2\pi)^3}{d{\bf 
p}_2\over(2\pi)^3}{d{\bf p}_3\over(2\pi)^3}(2\pi)^3\delta({\bf 
p}+{\bf p}_1 - {\bf p}_2 - {\bf p}_3)\nonumber\\
&\times &2\pi\delta(E+E_1-E_2-E_3) 
\left[(1+f)(1+f_1)f_2f_3-ff_1(1+f_2)(1+f_3)\right].
\label{eq:semi-classical limit7}\end{eqnarray}
In contrast, $C_{12}[f]$ describes collisions involving one 
condensate atom.
This kind of collision brings about equilibrium between the {\it 
excited} atoms and {\it condensate} atoms and it is given by (these 
collision terms were ignored in Ref.\cite{ref:Zargrinik})
\begin{eqnarray}
C_{12}[f]&=& 2g^2 \int{d{\bf p}_1\over(2\pi)^3}{d{\bf 
p}_2\over(2\pi)^3}{d{\bf p}_3\over(2\pi)^3}(2\pi)^3\delta({\bf p}_1 + 
{\bf p}_s - {\bf p}_2 - {\bf p}_3)\nonumber\\
& &\qquad\hbox{}\times 
2\pi\delta(E+\epsilon_c-E_2-E_3)(2\pi)^3[\delta({\bf p}-{\bf p}_1) 
-\delta({\bf p}-{\bf p}_2) - \delta({\bf p}-{\bf p}_3)]\nonumber\\
& &\qquad\hbox{}\times 
\left[n_c(1+f_1)f_2f_3-n_cf_1(1+f_2)(1+f_3)\right],
\label{eq:semi-classical limit8}
\end{eqnarray}
with $E_i({\bf r}, t) = {p_i^2\over 2m}+U({\bf r}, t)$ and $f_i 
\equiv f({\bf r, p}_i, t)$.
These two collision terms were first derived by Kirkpatrick and 
Dorfman \cite{ref:kirkdorf}, who considered a uniform gas and worked 
in a frame in which ${\bf v}_s({\bf r}, t) = 0$.  Our present 
calculation \cite{ref:Nikzargrif} is for trapped Bose gas and works 
in the lab frame.  As a result, (\ref{eq:semi-classical limit8}) 
takes into account that a condensate atom has energy $\epsilon_c 
\equiv \mu_c+{1\over 2} mv^2_s$ and momentum ${\bf p}_s = m{\bf 
v}_s$, where $\mu_c$ is defined in (\ref{eq:HF Popov equation 
motion6}).  We note that:
\begin{itemize}
\item $C_{22}[f]$ conserves number, momentum and energy of the 
colliding excited atoms and therefore
\begin{eqnarray} &\int& d{\bf p} C_{22}[f] = 0 \nonumber\\
&\int& d{\bf p}{\bf p} C_{22}[f] = 0 \nonumber\\
&\int& d{\bf p} E_p C_{22} [f] = 0 \label{eq:semi-classical 
limit9}\end{eqnarray}

\item $C_{12}[f]$ conserves momentum and energy of the colliding 
atoms, with the condensate 
atoms having energy $\epsilon_c$ and momentum $m{\bf v}_s$. 
Therefore, we have
\end{itemize}
\begin{eqnarray}  
&\int& d{\bf p} ({\bf p} - m{\bf v}_s) C_{12}[f] = 0 \nonumber\\
&\int& d{\bf p} (E_p - \epsilon_c) C_{12} [f] = 0, 
\label{eq:semi-classical limit10}\end{eqnarray}
but
\begin{equation} \int d{\bf p} C_{12}[f] \equiv \Gamma_{12}[f] \neq 0.
\label{eq:semi-classical limit11}\end{equation}
The fact that $\Gamma_{12}$ is finite follows since $C_{12}$ does 
{\it not} conserve the number of {\it excited} atoms; $\Gamma_{12}$ 
will be referred to as a source term.

As usual in dealing with the collision-dominated hydrodynamic region, 
we will assume that the collisions $C_{22}[f]$ produce {\it local 
thermal equilibrium} among the excited atoms.
This is described by the {\it local} Bose distribution:
\begin{equation}
f_0({\bf r, p}, t) = {1 \over e^{\beta[{({\bf p}-m{\bf v}_n)\over 
2m}^2+U({\bf r}, t) - {\tilde \mu}]}-1},
\label{eq:semi-classical limit12}
\end{equation}
where now $\beta, {\bf v}_n, U$ and ${\tilde \mu}$ all depend on 
${\bf r}, t$.  One can easily checks that $C_{22}[f_0]$ vanishes 
exactly.
This uses the key identity for Bose distribution 
given by (\ref{eq:Green function17}).  In contrast, one finds that 
substituting $f=f_0$ in 
(\ref{eq:semi-classical limit8}) gives
\begin{eqnarray}
C_{12} [f_0] &=& 2g^2 \int{d{\bf p}_1\over(2\pi)^3}{d{\bf 
p}_2\over(2\pi)^3}{d{\bf p}_3\over(2\pi)^3}(2\pi)^3\delta({\bf 
p}_1+{\bf p}_s - {\bf p}_2 - {\bf p}_3)\nonumber\\
&\times &2\pi\delta(E_1+\epsilon_c-E_2-E_3)(2\pi)^3[\delta({\bf 
p}-{\bf p}_1) -\delta({\bf p}-{\bf p}_2) - \delta({\bf p}-{\bf 
p}_3)]\nonumber\\
&\times &  \left[n_c-n_c e^{-\beta({\tilde\mu}' - \mu_c)}\right](1 + 
f_{01}) f_{02} f_{03},
\label{eq:semi-classical limit13}
\end{eqnarray}
where ${\tilde\mu}' \equiv{\tilde\mu} - {1\over 2}m ({\bf v}_n - {\bf 
v}_s)^2 \simeq{\tilde\mu}\ $ in a {\it linearized} theory.  We note 
that $C_{12}[f_0]$ in (\ref{eq:semi-classical limit13}) vanishes when 
\begin{equation} \mu_c ={\tilde\mu}, \label{eq:semi-classical 
limit14}\end{equation}
ie, only if the condensate atoms are in {\it diffusive equilibrium} 
with the non-condensate atoms.
The same kind of factor arises in the theory of the growth dynamics 
of a condensate in trapped Bose gases developed by Gardiner, Zoller 
and coworkers 
\cite{ref:Gardzoller1, ref:Gardzoller2}.

We next derive {\it hydrodynamic equations} for the non-condensate 
atoms by taking moments of the kinetic equation:
\begin{equation}
\left[{\partial\over\partial t}+{{\bf p}\over m}\cdot
\mbox{\boldmath$\nabla$}_r
- \mbox{\boldmath$\nabla$}U({\bf r}, 
t)\cdot\mbox{\boldmath$\nabla$}_p\right] f({\bf r, p}, t) = C_{22}[f] 
+ C_{12}[f].
\label{eq:semi-classical limit15}
\end{equation}
We obtain \cite{ref:Zargrinik, ref:Nikzargrif}
\begin{eqnarray}
\int d{\bf p} [\mbox{kinetic eq.}] &\rightarrow& {\partial{\tilde 
n}\over\partial t}+\mbox{\boldmath$\nabla$}\cdot{\tilde n}{\bf v}_n = 
\Gamma_{12}\nonumber\\
\int d{\bf pp} [\mbox{kinetic eq.}] &\rightarrow& m{\tilde 
n}\left({\partial\over\partial t}+{\bf 
v}_n\cdot\mbox{\boldmath$\nabla$}\right){\bf v}_n = 
-\mbox{\boldmath$\nabla$}{\tilde P}-{\tilde 
n}\mbox{\boldmath$\nabla$}U - m({\bf v}_n - {\bf 
v}_s)\Gamma_{12}\nonumber\\
\int d{\bf p}p^2 [\mbox{kinetic eq.}] &\rightarrow& {\partial {\tilde 
P}\over\partial t}+\mbox{\boldmath$\nabla$}\cdot({\tilde P}{\bf v}_n) 
= 
-{2\over 3}{\tilde P}\mbox{\boldmath$\nabla$}\cdot{\bf v}_n\nonumber\\
& \quad & \quad +{2\over 3}[\mu_c + {1\over 2}m({\bf v}_n - {\bf 
v}_s)^2 - U]\Gamma_{12}.
\label{eq:semi-classical limit16}
\end{eqnarray}
Here $\Gamma_{12}$ is defined in (\ref{eq:semi-classical limit11}) and
\begin{eqnarray}{\tilde n}({\bf r}, t)&\equiv& \int{d{\bf 
p}\over(2\pi)^3}f_0({\bf r, p,} t) = 
{1\over\Lambda^3}g_{3/2}(z)\nonumber\\
{\tilde P}({\bf r}, t) &\equiv& \int{d{\bf p}\over(2\pi)^3}{p^2\over 
3m}\left.f_0({\bf r, p,} 
t)\right|_{v_n=0}={1\over\beta}{1\over\Lambda^3}g_{5/2}(z), 
\label{eq:semi-classical limit17}
\end{eqnarray}
with the local fugacity $z({\bf r}, t)= e^{\beta({\tilde\mu}-U)}$.

\section{The two-fluid hydrodynamic equations}
\label{sec:The two-fluid hydrodynamic equations}

In Sections~\ref{sec:Hydrodynamic oscillations in a trapped Bose gas} 
and \ref{sec:Dynamics of the non-condensate atoms}, 
 we derived a set of 7 equations for non-condensate and the 4 
equations for the condensate.
This gives a coupled set of equations for the variables:
\begin{equation} n_c, {\tilde n}, {\bf v}_n, {\bf v}_s, {\tilde P}\ 
\mbox{and}\ \Gamma_{12} \equiv\int{d{\bf 
p}\over(2\pi)^3}C_{12}[f_0].\label{eq:two-fluid1}\end{equation}
>From (\ref{eq:semi-classical limit13}), one sees $\Gamma_{12}$ is 
function of $f_0$, which in turn depends on $\beta, {\bf v}_n, 
{\tilde n}, n_c$ and ${\tilde \mu}$. Thus our system of hydrodynamic 
equations is closed and can be solved since it involves 11 local 
variables and 11 equations.

For reasons discussed earlier, the explicit effect of the $C_{22}$ 
collisions has disappeared in the hydrodynamic equations 
(\ref{eq:semi-classical limit16}).
However, the $C_{22}$ collisions justify and enforce the local 
equilibrium form 
(\ref{eq:semi-classical limit12}) 
used for $f({\bf r, p}, t), $ which describes the excited atoms.
In contrast, $C_{12}$ is still explicitly present in 
(\ref{eq:semi-classical limit16})
through the $\Gamma_{12}$ source term.
This means that the condensate is not in {\it diffusive} equilibrium 
with 
non-condensate.
We recall from 
(\ref{eq:semi-classical limit13}) 
that 
\begin{equation} \Gamma_{12}\propto g^2 
n_c[1-e^{-\beta({\tilde\mu}-\mu_c)}], 
\label{eq:two-fluid2}\end{equation}
and thus $\Gamma_{12}[f_0] $ vanishes only when ${\tilde\mu}=\mu_c$.  
One might expect that this condition would arise in the case of 
strong $C_{12}$ collisions.  A detailed study \cite{ref:Nikzargrif}
 shows that even in this case, one must allow for fluctuations in the 
variable $\mu_{diff}\equiv{\tilde\mu}-\mu_c$ and thus we cannot 
simply set $\Gamma_{12} = 0.$ Since $\Gamma_{12}\neq 0$ in 
(\ref{eq:semi-classical limit16}), we have to generalize the HF Popov 
description to ensure that [compare with the result in (\ref{eq:HF 
Popov equation motion5})] 
\begin{equation} {\partial n_c\over\partial 
t}+\mbox{\boldmath$\nabla$}\cdot(n_c{\bf v}_s) = 
-\Gamma_{12}.\label{eq:two-fluid3}\end{equation}
This is needed for the continuity equation to be correct
\begin{equation}{\partial n_c\over\partial 
t}+\mbox{\boldmath$\nabla$}\cdot{\bf j}=0, 
\label{eq:two-fluid4}\end{equation}
where
\begin{eqnarray} n&=&{\tilde n}+n_c\nonumber\\
j&=&{\tilde n}{\bf v}_n+n_c{\bf 
v}_s.\label{eq:two-fluid5}\end{eqnarray}
As discussed in detail in Ref.\cite{ref:Nikzargrif}, a simple 
microscopic model based on the neglect of anomalous pair correlation 
functions $\langle{\tilde\psi}({\bf r}){\tilde\psi}({\bf r})\rangle$ 
leads to both (\ref{eq:two-fluid3}) and (\ref{eq:semi-classical 
limit16}).  

We can linearize our coupled two-fluid hydrodynamic equations given 
by (\ref{eq:HF Popov equation motion5}) and (\ref{eq:semi-classical 
limit16}) around the static equilibrium solution, where $\mu_{c0} = 
{\tilde\mu}_0$. 
The fluctuations $\delta{\tilde n}, \delta n_c, \delta{\bf v}_n, 
\delta{\bf v}_s$ are precisely the quantities which can be directly 
measured in trapped gases.
The only static equilibrium functions we need are ${\tilde n}_0, 
n_{c0}$ and ${\tilde P}_0$, which are given by 
(\ref{eq:semi-classical limit17}),
\begin{equation}
{\tilde P}_0 = {1\over\beta_0\Lambda^3_0}g_{5/2} (z_0), \ {\tilde 
n}_0 ={1\over\Lambda^3_0}g_{3/2} (z_0), 
\label{eq:two-fluid6}
\end{equation}
where $z_0 = e^{-\beta gn_{c0}({\bf r})}$ and $ n_{c0}$ is given by 
the {\it static} HFP equation (\ref{eq:Beyond6}) for $\Phi({\bf 
r})$.  We note that
these static thermodynamic properties for a Bose-condensed gas are 
quite different from the expressions first given by Lee and Yang 
(1958) for a uniform gas \cite{ref:Leeyang}.
They calculated only the first-order interaction corrections to the 
thermodynamic properties of an ideal Bose gas, using
\begin{eqnarray} {\tilde P}_0 &\simeq& {1\over\beta_0\Lambda^3_0} 
g_{5/2} (z=1) - gn_{c0}{\tilde n}_{cr}\nonumber \\
{\tilde n}_0 &\simeq & {1\over\Lambda^3_0} g_{3/2} (z=1) \equiv 
{\tilde n}_{cr}.
\label{eq:two-fluid7}\end{eqnarray}
Here ${\tilde n}_{cr}$ is the critical density at the temperature $T$ 
for an ideal uniform Bose gas.  The non-perturbative approximation we 
use for thermodynamic properties [as illustrated 
by (\ref{eq:two-fluid6})] is consistent with the Lee-Yang results 
when we expand to first order in $g$ \cite{ref:Grifzar}.

We can use our generalized ZGN equations to derive two-fluid 
equations in the traditional form first given by  Landau (1941).
In linearized form, these are 
\cite{ref:Land,ref:Khalatnikov}
\begin{eqnarray}
{\partial\delta n\over\partial t} &+& 
\mbox{\boldmath$\nabla$}\cdot\delta{\bf j} = 0\nonumber\\
m {\partial\delta {\bf j}\over\partial t} &=& 
-\mbox{\boldmath$\nabla$}\delta P-\delta 
n\mbox{\boldmath$\nabla$}U_{\rm ext}({\bf r})\nonumber\\
m{\partial\delta {\bf v}_s\over\partial t} &=& 
-\mbox{\boldmath$\nabla$}\delta\mu_c\nonumber\\
{\partial\delta s\over\partial t} &=& 
-\mbox{\boldmath$\nabla$}\cdot(s\delta{\bf 
v}_n),\label{eq:two-fluid8}\end{eqnarray}
where we have included an external potential, $s({\bf r}, t)$ is the 
local entropy density and 
\begin{eqnarray} \delta{\bf j} &=& n_{c0}\delta{\bf v}_s + {\tilde 
n}_0\delta{\bf v}_n\nonumber\\
\delta n&=&\delta n_c+\delta{\tilde 
n}.\label{eq:two-fluid9}\end{eqnarray}
Of course, the Landau equations describe the case when the superfluid 
and normal fluid are in dynamic local equilibrium with each other.
The Landau form of two-fluid equations is very natural for superfluid 
$^4$He, for which they were developed.
They involve the usual thermodynamic variables: pressure,  entropy 
and temperature fluctuations.
However, in a trapped gas, the more natural variables are those used 
in a ZGN-type formulation 
\cite{ref:Zargrinik,ref:Nikzargrif},
namely ${\tilde n},  n_c, {\bf v}_n, {\bf v}_s$ and $\mu_{diff}$.

In the generalized ZGN equations discussed in Ref. 
\cite{ref:Nikzargrif}, the fluctuations around static equilibrium 
involve fluctuations in $\delta\Gamma_{12}$, namely [see 
(\ref{eq:two-fluid2})]
\begin{equation}
\delta\Gamma_{12}\propto g^2 
n_{c0}\delta\mu_{diff.}\label{eq:two-fluid10}
\end{equation}
This leads to an additional equation of motion for 
$\delta\mu_{diff}$, with a characteristic relaxation time 
$\tau_\mu$.  This gives rise to a new relaxational mode of the
condensate. The ZGN limit \cite{ref:Zargrinik} corresponds to 
$\omega\tau_\mu\gg 1$ while the Landau two-fluid limit corresponds to 
$\omega\tau_\mu\ll 1.$  While the relaxational mode is not included, 
the ZGN equations appear to give a reasonable first approximation for 
the other hydrodynamic modes, as shown in the explicit results given 
in Ref.\cite{ref:Grifzar}. 
We note that Landau two-fluid equations given by 
(\ref{eq:two-fluid8}) have been recently used by Shenoy and Ho 
\cite{ref:Shenoyho} to work out the hydrodynamic normal modes of a 
trapped superfluid gas.

The Landau two-fluid equations, which assume total local equilibrium, 
are valid at all temperatures.
Our present analysis is restricted to finite $T\sim T_{\rm BEC}$, but 
could be generalized to the case of very low temperatures.
At $T=0$, of course, the distinction between the collisionless and 
hydrodynamic region is not a useful one (see, moreover, the comment 
at the end of Section~\ref{sec:Why are excitations in a Bose fluid so 
interesting}).  However, at $T=0$, we have $\rho_n=0$ and 
$\rho_s=\rho$ and then the Landau two-fluid equations formally reduce 
to hydrodynamic equations for {\it one} component:
\begin{eqnarray} & &{\partial n\over\partial t} +\nabla\cdot n{\bf v} 
= 0\nonumber\\
& &m{\partial{\bf v}\over\partial t} = - 
\mbox{\boldmath$\nabla$}\mu_c.\label{eq:two-fluid11}
\end{eqnarray}
These equations are the basis of recent work by Pitaevskii and 
Stringari \cite{ref:Pitaevskii1}.  In conjuction with the local 
density approximation \cite{ref:Dalgiopitstr}
\begin{equation} \mu_c({\bf r}, t) =\mu_0[n({\bf r}, t)] + U_{\rm 
ext}({\bf r}), 
\label{eq:two-fluid12}\end{equation}
where $\mu_0$ is the chemical potential of a {\it uniform} 
interacting Bose-condensed gas at $T=0$, such as given by 
(\ref{eq:approximation1}), one may use (\ref{eq:two-fluid11}) to find 
collective mode frequencies past the MFA (or GP approxiamation).

\section{In-phase oscillation of the condensate and non-condensate}
\label{sec:In-phase oscillation of the condensate and non-condensate}

Two interesting collective modes exhibited by the ZGN equations 
involve 
{\it rigid} centre-of-mass oscillations of the 
equilibrium density profiles:
\begin{eqnarray}
\tilde n({\bf r}, t) 
&=& \tilde n_0({\bf r} - {\mbox{\boldmath$\eta$}}_n(t)), 
\quad
{\dot{\mbox{\boldmath$\eta$}}}_n(t) = {\bf v}_n(t) \nonumber \\
n_c({\bf r}, t) 
&=&  n_{c0}({\bf r} - {\mbox{\boldmath$\eta$}}_c(t)), 
\quad 
\dot{\mbox{\boldmath$\eta$}}_c(t) = {\bf v}_s(t).
\label{eq:oscillation of the cond1}
\end{eqnarray}
One finds \cite{ref:Zargrinik} an in-phase solution (Kohn mode) with 
${\bf v}_n = {\bf v}_s$, with a frequency given by the trap frequency 
$\omega_\alpha$; and an out-of-phase solution, with ${\bf v}_n$ and 
${\bf v}_s$ in opposite directions.
The latter  is the {\it analogue} of the second sound mode in liquid 
$^4$He.

First let us work out the static equilibrium solution $f_{eq}({\bf r, 
p})$ of 
(\ref{eq:semi-classical limit15}), 
which must satisfy
\begin{equation} 
{{\bf p}\over m}\cdot\mbox{\boldmath$\nabla$}_r f_{eq}({\bf r, p}) 
-\mbox{\boldmath$\nabla$}U_0({\bf r})\cdot\nabla_p f_{eq}({\bf r, 
p})= C[f_{eq}] = 0.
\label{eq:oscillation of the cond2}
\end{equation}
The solution is easily verified to be
\begin{equation} f_0({\bf r, p}) =  {1 \over e^{\beta_0({p^2\over 
2m}+U_0(r) - {\tilde\mu}_0)}-1}, \label{eq:oscillation of the 
cond3}\end{equation}
where (within the Thomas-Fermi approximation for $\mu_{c0}$)
\begin{eqnarray}
{\tilde\mu}_0&=& \mu_{c0}=U_{\rm ext}({\bf r})+2g{\tilde n}_0({\bf 
r})+gn_{c0}({\bf r})\nonumber\\
U_0({\bf r}) &=& U_{\rm ext}({\bf r})+2g[{\tilde n}_0({\bf r}) + 
n_{c0}({\bf r})].
\label{eq:oscillation of the cond4}
\end{eqnarray}
Using (\ref{eq:oscillation of the cond3}), 
we find ${\tilde n}_0({\bf r})$ is given by (\ref{eq:two-fluid6}), 
with
\begin{equation}
\Lambda_0 = \left({2\pi\over mk_B T_0}\right)^{1/2}; z_0\equiv 
e^{\beta_0({\tilde\mu}_0-U_0)}=e^{-\beta_0gn_{c0}}.\label{eq:oscillation 
of the cond5}
\end{equation}

We next consider the in-phase solution described by 
(\ref{eq:oscillation of the cond1}), with 
${\bf v}_s = {\bf v}_n = {\dot{\mbox{\boldmath$\eta$}}}$ 
(not dependent on the position ${\bf r}$). This is a solution of ZGN 
equations with 
$\eta_\alpha(t) \sim e^{i\omega_\alpha t}$,
where $\omega_\alpha$ is the parabolic trap frequency in 
$\alpha^{th}$ direction $(x, y, z)$.
The proof proceeds as follows (in the linearized theory).
Given (\ref{eq:oscillation of the cond1}), we have
\begin{eqnarray}
\delta n_c({\bf r}, t) 
&=& -{\mbox{\boldmath$\eta$}}(t) \cdot 
\mbox{\boldmath$\nabla$}n_{0c}({\bf r}) \nonumber \\ 
\delta{\tilde n}({\bf r}, t) 
&=& -{\mbox{\boldmath$\eta$}}(t) \cdot 
\mbox{\boldmath$\nabla$}{\tilde n}_0({\bf r}),
\label{eq:oscillation of the cond6}\end{eqnarray}
and hence
\begin{eqnarray} 
m \frac{\partial\delta{\bf v}_s}{\partial t} = 
-\mbox{\boldmath$\nabla$}\delta\mu_c 
&=& -\mbox{\boldmath$\nabla$}[2g\, \delta \tilde n + g\, \delta n_c] 
\nonumber \\
&=& \mbox{\boldmath$\nabla$}
\left({\mbox{\boldmath$\eta$}} \cdot [2g 
\mbox{\boldmath$\nabla$}{\tilde n}_0 + g 
\mbox{\boldmath$\nabla$}n_{c0}] \right).\label{eq:oscillation of the 
cond7}
\end{eqnarray}
This result is equivalent to [using (\ref{eq:oscillation of the 
cond4})]
\begin{equation}  
m \frac{\partial^2{\mbox{\boldmath$\eta$}}}{\partial t^2} = - 
\mbox{\boldmath$\nabla$} \left({\mbox{\boldmath$\eta$}} \cdot 
\mbox{\boldmath$\nabla$}U_{\rm ext} \right),
\label{eq:oscillation of the cond8}
\end{equation}
with 
\begin{equation} 
U_{\rm ext} = \frac{1}{2} m (\omega_x^2 x^2 + \omega_y^2 y^2 + 
\omega_z^2 z^2).
\label{eq:oscillation of the cond9}\end{equation}
The solution of 
(\ref{eq:oscillation of the cond8}) 
is clearly
\begin{equation} 
\ddot\eta_\alpha = -\omega_\alpha^2\eta_\alpha,
\quad
\eta_\alpha(t) \sim e^{i\omega_\alpha t}.
\label{eq:oscillation of the cond10}\end{equation}

In a similar way, one can show that the hydrodynamic equations for 
the non-condensate lead to the same equation of 
motion (\ref{eq:oscillation of the cond8}).
The rigid motion of the 
non-condensate density profile corresponds to a distribution function 
given by
\begin{eqnarray} 
f({\bf r, p}, t) 
&=& f_0({\bf r} - {\mbox{\boldmath$\eta$}}, {\bf p} - m 
{\dot{\mbox{\boldmath$\eta$}}}).
\label{eq:oscillation of the cond11}
\end{eqnarray}
This can be verified to solve the kinetic equation 
(\ref{eq:semi-classical limit15}), using the fact that $C[f_0] = 0$, 
and leads to the expression in (\ref{eq:oscillation of the cond1}), 
namely
\begin{equation}
\tilde n({\bf r}, t) 
= \int {d{\bf p}\over(2\pi)^3} f({\bf r}, {\bf p}, t) 
= \tilde n_0 ({\bf r} - {\mbox{\boldmath$\eta$}}).
\end{equation}
This in-phase dipole mode was first exhibited by ZGN in the 
hydrodynamic region \cite{ref:Zargrinik}. 
It is an important test of the proper treatment of the non-condensate 
dynamics.
This kind of mode is generic, existing in the collisionless region as 
well \cite{ref:Mintosi, ref:Bijlsmastoof}.

\section{First and second sound in a uniform Bose-condensed gas}
\label{sec:First and second sound in a uniform Bose-condensed gas}

The simplicity of the ZGN two-fluid equations allow a very 
transparent discussion of first and second sound in a uniform gas 
\cite{ref:Grifzar}.
We find two sound-wave solutions: 
$\omega = u_1 q$,
$\omega = u_2q$.
First sound has a velocity given by (to lowest order in $g$)
\begin{equation} 
u_1^2 = \frac{5}{3}\frac{\tilde P_0}{m \tilde n_0} 
+ \frac{2g \tilde n_0}{m}, 
\label{eq:sound velocity1}
\end{equation}
and mainly involves the non-condensate, 
with $v_n \gg v_s$ (but in-phase).
Second sound has a velocity given by (to lowest order in $g$)
\begin{equation} 
u^2_2 = \frac{gn_{c0}}{m}, \label{eq:sound velocity2}
 \end{equation}
and mainly involves the oscillation of the condensate, 
with $v_s\gg v_n$ (and out-of-phase).
It is clear that second sound is the collision-dominated hydrodynamic 
mode which is the 
{\it analogue} of the collisionless Bogoliubov phonon 
(\ref{eq:Green function13}) 
which we discussed in Section~\ref{sec:Green's function 
formulation}.  The frequency vanishes (becomes ``soft'') above 
$T_{\rm BEC}$ and it is the expected Goldstone mode of the Bose 
broken-symmetry \cite{ref:Bog2, ref:Hohmar}.  
In contrast to liquid $^4$He, both second and first sound in a Bose 
gas involve density fluctuations (ie, second sound is not a 
temperature wave in a gas!).

Can we study first and second sound pulse propagation in cigar-shaped 
traps?  
In this volume, Ketterlee \cite{ref:Ketterle} reviews the beautiful 
pulse propagation experiments carried out at MIT at low temperatures 
in the collisionless region \cite{ref:Andkurnmiedurftowninket}.  
If we could get into the hydrodynamic region, 
one should be able to see {\it two} pulses propagate---roughly with 
the speeds $u_1$ and $u_2$ given above.
This would be very dramatic and a direct confirmation of the 
superfluid dynamics described by the two-fluid equations.
Inside the broad, relatively uniform condensate distribution along 
the axis of the 
cigar-shaped trap, 
one could excite both first and second sound pulses, 
with the relative intensity of each propagating pulse given by 
\cite{ref:Nikgrif1}
\begin{equation} 
\frac{1}{u^2_i} \delta(z - u_it).\label{eq:sound velocity3}
\end{equation}
The velocity of second sound would give [using (\ref{eq:sound 
velocity2})] a direct measurement of the condensate density $n_{c0}$.
Perhaps of more interest, the velocity of the first sound pulse would 
give a relatively 
{\it direct} measurement [using (\ref{eq:sound velocity1})] of the 
non-condensate density ${\tilde n}_0$ ``underneath'' the condensate.

\section{Concluding remarks}
\label{sec:Concluding remarks}

In Sections \ref{sec:Dynamics of the non-condensate atoms} and 
\ref{sec:The two-fluid hydrodynamic equations}, we discussed the 
linearized form of the generalized two-fluid equations 
\cite{ref:Nikzargrif} which describe how the condensate relaxes to 
equilibrium with the excited atoms. These hydrodynamic equations with 
$\Gamma_{12} \neq 0$ 
(i.e.
$\tilde\mu \neq \mu_c$) 
should also be useful in the study of how a condensate grows and 
comes into equilibrium with the thermal cloud \cite{ref:Ketterle}.
While based on a different formalism, our theory has points of 
contact with the work by Gardiner, Zoller and co-workers on the 
kinetics of condensate growth 
\cite{ref:Gardzoller1,ref:Gardzoller2}.  However, our analysis, while 
assuming the existence of a condensate, takes the dynamical evolution 
of the non-condensate fully into account.  The two-fluid dynamics of 
trapped Bose-condensed gases promises to be a very rich subject, as 
discussed in Ref. \cite{ref:Nikzargrif} .
The region where $\Gamma_{12} \neq 0$ 
is not easily studied in superfluid $^4$He, 
largely because local thermal equilibrium is almost too easy to reach 
in a liquid.
However, some years ago Pitaevskii \cite{ref:Pitaevskii2} treated the 
regime very close to 
$T_\lambda$,
when the superfluid density $\rho_s({\bf r}, t)$ was not necessarily  
 in equilibrium with the normal fluid density $\rho_n({\bf r}, t)$.  

One of the interesting aspects of trapped Bose gases is that even if 
the centre of the trap is at large enough density so that one is in 
the hydrodynamic region described by dynamic local thermal 
equilibrium, this description will always break down in the low 
density non-condensate thermal cloud far enough from the trap centre.
Recent theoretical studies 
\cite{ref:Nikgrif2,ref:Kavpethsmith}
have shown that the cross-over region between the hydrodynamic and 
collisionless regions in this low density region plays the dominant 
role as the source of damping of certain kinds of hydrodynamic 
oscillations.

More generally, I would like to emphasize that the study of 
collective modes in trapped Bose gases finally provides us with the 
opportunity of making a quantitative test of the complex dynamics of 
a system with a Bose-broken symmetry.
As I discuss elsewhere in this volume \cite{ref:Grif1}, the 
complexity of dealing with a liquid such as superfluid $^4$He never 
allowed one to make a thorough confrontation of the microscopic 
theory of Bose-condensed systems with experiment \cite{ref:Grif2}.
Bose gases finally allow us to do this, in a much cleaner fashion 
\cite{ref:Ketterle, ref:Stamper}.
I would hope that future studies on trapped Bose gases will also 
stimulate new interest in the dynamics of superfluid $^4$He as a 
Bose-condensed liquid.

\acknowledgments
The work presented here on the two-fluid hydrodynamic equations has 
been done in close collaboration with Eugene Zaremba and Tetsuro 
Nikuni.
In addition, I have benefitted from stimulating discussions with many 
of my colleagues, and especially would like to thank K.Burnett, M.L. 
Chiofalo, E. Cornell, W. Ketterle, L.P.
Pitaevskii, G.V. Shlyapnikov, D.M. Stamper-Kurn, S.Stringari and P. 
Zoller.

These lectures are partly based on research done while participating 
in the 1998 BEC Workshop sponsored by the Institute of Theoretical 
Physics in Santa Barbara.

My research is supported by NSERC of Canada.

\newpage

\centerline{\bf Figure Captions}
\begin{enumerate}
\item Hartree-Fock self-energies in a normal Bose gas.

\item Various interaction terms involving condensate (wiggly line) 
and non-condensate (solid line) atoms.

\item Self-energy diagrams in the Bogoliubov approximation $(T = 0)$.

\item Self-energy diagrams in the first-order Popov approximation.  
The propagators are for an ideal Bose gas.

\item Self-consistent Hartree-Fock-Bogoliubov (HFB) approximation for 
self-energies.

\item Ladder diagram approximation for the $t$-matrix.
\end{enumerate}

\end{document}